\documentclass[12pt]{article}
\usepackage[latin1]{inputenc}
\usepackage[numbers]{natbib}
\usepackage[a4paper,left=2.5cm,right=2.5cm,top=2cm,bottom=1.5cm,footnotesep=.75cm,headheight=13.6pt]{geometry}
\usepackage{setspace}
\usepackage{amsmath}
\usepackage{graphicx}
\usepackage[small]{caption}
\usepackage{authblk}
\usepackage{rotating}
\usepackage{float}
\usepackage{pdflscape}
\usepackage{subcaption}
\usepackage{xcolor}
\usepackage{endnotes}
\usepackage{amssymb}

\title{Interpreting (and testing) factors loadings}
\author[1]{\large Antonio Monta\~{n}\'es}
\author[2]{\large Esther Ruiz\thanks{Corresponding author. E-mail: ortega@est-econ.uc3m.es\\ \textbf{Acknowledgements}: Financial support from the Spanish National Research Agency (Ministry of Science and Technology) Projects PID2023-150095NB-C44 (first  author) and PID2022-139614NB-C22 (second author), both funded by MCIN/AEI/10.13039/501100011033/FEDER, EU, is gratefully acknowledged. We are thankful for their comments to Matteo Barigozzi and rest of participants at the "Score-driven and nonlinear time series models, with applications in economics, finance and climate change" (Venice, May 2026). Data are available from the authors upon request. No conflict of interest is declared by the authors.}}
\affil[1]{Department of Economic Analysis, Universidad de Zaragoza (Spain)}
\affil[2]{Department of Statistics, Universidad Carlos III de Madrid (Spain)}

\date{\today}

\begin{document}

\maketitle

\begin{abstract}

Dynamic Factor Models (DFMs) are popular to reduce dimensionality being customary in the empirical analysis of large systems of macroeconomic and/or financial variables. In this context, the common underlying factors and their loadings are often extracted using Principal Components (PC), which are consistent and asymptotically normal under very general conditions. Consequently, inference on the factor loadings, which is crucial for the correct interpretation of the underlying factors, is often based on their asymptotic distribution with the limit covariance matrix of the loadings consistently estimated using HAC estimators. In this paper, we analyse the performance of the finite sample asymptotic approximation when constructing confidence intervals and testing about estimated PC loadings. We show that this approximation is seriously affected when the cross-sectional dimension is not large enough. We propose using HAR inference and a subsampling procedure to correct the MSE of the loadings to take into account the uncertainty associated with the estimation of the covariance matrix and of the factors, respectively. The relevance of the results is illustrated in an empirical analysis of economic convergence among the US states.

\end{abstract}




\setcounter{page}{1}

\newpage

\doublespacing

\section{Introduction}

Dynamic factor models (DFMs) are a popular tool to extract latent common components that summarize the information contained in large systems of economic and/or financial variables. Although several alternative factor extraction procedures are available in the related literature, Principal Components (PC) is still most popular in empirical applications due to its simplicity and good properties; see, for example, the recent description and comparison of alternative factor extraction procedures by Ruiz and Poncela (2022) and Barigozzi, Fresoli and Ruiz (2026). The little structure assumed by PC allows it to handle factor extraction in large systems of variables at small computational cost. Furthermore, under very general conditions, PC factors and loadings are consistent and asymptotically normal. Finally, note that even if the factors are extracted using the EM algorithm, PC is often used to obtain initial estimates of the factors; see Doz, Giannone and Reichlin (2012).

%The performance of the asymptotic distribution to approximate the finite sample distribution of PC factors, and therefore, to carry out inference on the estimated factors has been estudied in several papers; see, for example, Fresoli, Poncela and Ruiz (2023, 2024), Barigozzi, Fresoli and Ruiz (2026) and Bellocca \textit{et al}. (20026). However, as far as we are concerned, there is not any study of the properties of the asymptotic distribution when approximatingthe finite sample distribution of PC loadings.
Inference on PC factors and loadings, which is crucial for their adequate interpretation in empirical applications, is often based on their respective asymptotic distributions derived by Bai (2003). Several works analyse the properties of the asymptotic distribution to approximate the finite sample distribution of PC factors. These properties depend on two main elements. First, it depends on the cross-sectional and temporal sample sizes, and their proportion. Second, they depend on whether the idiosyncratic components are cross-sectionally homoscedastic and/or correlated. %Third, the presence of factors that only load on particular variables within the system should also be considered when analysing the finite sample performance of PC factors. 
Among those who analyse the finite sample performance of PC factors, an early contribution is Boivin and Ng (2006), who, in the context of using PC factors to summarize the information contained in a large number of potential predictors, study the role of these elements on forecast performance. They find evidence about the cross-sectional dimension of the system not needing to be extremely large for PC factors to be reasonably precise; see also Poncela and Ruiz (2016) for the same conclusion. %Furthermore, they also conclude that, in the presence of idiosyncratic cross-correlation, increasing the cross-sectional dimension may lead to extracted factors with less predictive power; see also 
More recently, Fresoli, Poncela and Ruiz (2023, 2024) and Bellocca \textit{et al}. (2026) study the effects of idiosyncratic cross-correlation on the asymptotic approximation of the covariance matrix of PC factors with the latter authors considering also the effect of idiosyncratic cross-sectional heteroscedasticity. Finally, Barigozzi, Fresoli and Ruiz (2026) analyse the asymptotic distribution of PC factors under different structures of the idiosyncratic covariance matrix and compare it with that of the factors extracted using linear projections or the Kalman filter.

Drawing inference on factor loadings is also crucial for the empirical interpretation of the factors. Practitioners and academics often focus on the largest loadings to interpret the factors; see, for example, the discussions by Cadima and Jolliffe (1995) or, more recently, by Lettau and Pelger (2020), Pelger (2020), Pelger and Xiong (2022), Despois and Doz (2023), Daniele and Schnaitmann (2026), Freyaldenhoven (2026) and Wei and Zhang (2026). In empirical applications, Lehmann and Modest (1988) use the loadings to construct portfolios, while they are used to interpret economic global and geographic-specific factors of the euro area by Dias, Pinheiro and Rua (2015) and of temperatures in the US by Gospodinov, Lopez Gaffney and Ng (2025). Caporin, Rodr\'iguez-Caballero and Ruiz (2024) determine the factor structure of a DFM based on the analysis of the loadings, and Gonz\'alez-Alvarez \textit{et al}. (2024) use the loadings to identify clubs of convergence among Balkan economies.\endnote{Loadings have also been used to interpret factors in order areas as, for example, ecological data; see Jackson (1993) and Peres-Neto, Jackson and Somers (2003), among others.} 

However, as far as we know, the analysis of the performance of the asymptotic distribution of PC factor loadings to approximate their finite sample distribution is not available. The first contribution of this paper is to fill this gap and analyse the role of the cross-sectional and temporal sample sizes, and of the properties of the idiosyncratic components on the performance of confidence intervals and significance tests on PC loadings when they are constructed based on their asymptotic distribution. We carry out extensive Monte Carlo simulations using as data generating processes (DGPs) models with cross-sectional and temporal dimensions close to those often encountered in empirical applications. We also consider different structures of the DFM with weakly serial and cross-correlated and/or heteroscedastic idiosyncratic components. We show that the performance of the significance tests and confidence intervals for PC loadings is seriously affected when the covariance matrix of the loadings is estimated assuming that the idiosyncratic components are white noise. Using the Bartlett kernel to estimate the covariance matrix of the loadings taking into account serially idiosyncratic dependence, as proposed by Bai (2003), helps but inference still has poor properties and the interpretation of the factors can be seriously  blurred. The second contribution of this paper is the proposal of a more efficient estimator of the covariance of the loadings based on the QS kernel instead of the Bartlett kernel, with the number of lags chosen as proposed within the Heteroscedasticity and Autocorrelation Robust (HAR) framework. We also propose to correct the estimated covariance matrix using subsampling to take into account the uncertainty associated with the estimation of the factors, which can be important when the cross-sectional dimension is not large enough and the factors are poorly estimated. Finally, as it is often done in the context of regression with autocorrelated noises, we propose using fixed-b critical values instead of the normal critical values (HAR inference) to take into account the fact that in practice, inference is carried out using an estimated covariance instead of the true one. The proposed methodology delivers confidence intervals and significance tests for the loadings with coverages and sizes closer to the respective nominal levels. 

Our final contribution is the illustration of the results in an empirical application to study the convergence of \textit{per capita} GDP among the US states. When the analysis of significance of the loadings is carried out using tests computed with the covariance matrix of the loadings estimated assuming serially uncorrelated idiosyncratic components and normal critical values, the number of significant loadings is large and consequently, the interpretation of the factors is diffuse, mixing factors that affect all states with those that load only on particular groups of states. However, computing the covariance matrix as proposed in this paper with inference being based on the HAR critical values, allows to separate more neatly the factors loading on all states from those loadings on groups of states. In doing so, we allow for the identification of some coherent economic regions: i) an occidental/metropolitan region as opposite to region with more traditional production; ii) the Sun Belt/Southeast region as opposite to the Plains-Midwest-North-east region; iii) the manufacturing Midwest region as opposite to states with natural resources and energy; and iv) regions associate with trading channels.

The rest of this paper is organized as follows. Section \ref{sec:DFM} describes the approximate DFM, PC estimation and the asymptotic distribution of PC loadings. Section \ref{sec:testing} describes the proposed feasible tests and confidence intervals for PC loadings. Section \ref{sec:MC} is devoted to the results of the Monte Carlo simulations designed to analyse the coverage of point-wise confidence intervals for PC loadings and the size and power of individual significance tests. Section \ref{sec:empirics} illustrates the results with the empirical analysis of convergence of \textit{per capita} GDP in the US states. Finally, Section \ref{sec:conclusions} summarizes the main conclusions.

\section{Approximate Dynamic factor model and Principal Components}
\label{sec:DFM}

In this section we describe the approximate DFM, PC estimation and the asymptotic distribution of PC loadings.

\subsection{The DFM and PC estimation}

Consider the $N \times 1$ vector of stationary zero mean time series $\mathbf{Y}_t=\left(y_{1t},...,y_{Nt} \right) ^{\prime}$, observed from $t=1,...,T$, which can be represented by the following DFM
\begin{equation}
\label{eq:DFM}
\mathbf{Y}_t=\mathbf{\Lambda F}_t + \boldsymbol{\varepsilon}_t,
\end{equation}
where $\mathbf{\Lambda}:=\left( \boldsymbol{\lambda}_1,...,\boldsymbol{\lambda}_N \right)^{\prime}$ is the $N\times r$ matrix of factor loadings, with $\boldsymbol{\lambda}_i$ being its $i$'th row for $i=1,...,N$, $\mathbf{F}_t= \left(F_{1t},...,F_{rt} \right)^{\prime}$ is the $r \times 1$ vector of underlying unobserved factors, and $\boldsymbol{\varepsilon}_t=\left(\varepsilon_{1t},...,\varepsilon_{Nt} \right)^{\prime}$ is the $N \times 1$ vector of idiosyncratic components with zero mean and contemporaneous covariance matrix $\mathbf{\Sigma}_{\varepsilon}$. We call $\mathbf{C}_t:=\mathbf{\Lambda} \mathbf{F}_t$ the $N \times 1$ vector of common components. Furthermore, let $\mathbf{Y}:= \left( \mathbf{Y}_1,...,\mathbf{Y}_T \right)^{\prime}$ be the $T \times N$ matrix of observations and $\mathbf{F}:= \left( \mathbf{F}_1,...,\mathbf{F}_T\right)^{\prime}$ be the $T \times r$ matrix of factors. Note that although the relation between the factors and the observable variables is static, model (\ref{eq:DFM}) is popularly known as "Dynamic" because the factors may be dynamic with temporal dependence. As it is common in this literature, the number of factors, $r<N$, is assumed to be known and fixed; see Barigozzi and Hallin (2026) for a recent description of fundamental issues in the theory and practice of factor models.

The following assumptions are made about the loadings, factors and idiosyncratic components in the DFM in (\ref{eq:DFM}):

\begin{enumerate}

\item Common components (Loadings and factors)%, which are assumed to be fixed unknown constants

\begin{enumerate}
\item $\lim_{N\rightarrow\infty} \parallel \frac{\mathbf{\Lambda}^{\prime} \mathbf{\Lambda}}{N}-\mathbf{\Sigma}_{\Lambda}\parallel =0$, where $\mathbf{\Sigma}_{\Lambda}$ is an $r \times r$ finite positive definite matrix, and, for all $i \in  \mathbb{N}$, $\parallel \boldsymbol{\lambda}_i \parallel \le M_{\Lambda}$ for some finite positive $M_{\Lambda}$ independent of $i$.

\item $\lim_{N\rightarrow\infty} \parallel \frac{\mathbf{\Lambda}^{\prime} \mathbf{\Sigma}_{\varepsilon}^{-1} \mathbf{\Lambda}}{N}-\mathbf{\Sigma}^{*}_{\Lambda}\parallel =0$, where $\mathbf{\Sigma}^{*}_{\Lambda}$ is an $r \times r$ positive definite matrix.

\item $\lim_{N\rightarrow\infty} \parallel \frac{\mathbf{\Lambda}^{\prime} \mathbf{\Sigma}_{\varepsilon} \mathbf{\Lambda}}{N}-\mathbf{\Sigma}^{\dagger}_{\Lambda}\parallel =0$, where $\mathbf{\Sigma}^{\dagger}_{\Lambda}$ is an $r \times r$ positive definite matrix.

\item $\forall t, \parallel \mathbf{F}_t \parallel \leq c$, where $c$ is a large enough constant. 

\item $\lim_{T\rightarrow\infty} \frac{1}{T} \sum_{t=h+1}^T \left( \mathbf{F}_t-\bar{\mathbf{F}} \right) \left( \mathbf{F}_{t-h} - \bar{\mathbf{F}} \right) ^{\prime} = \mathbf{\Sigma}_{F}(h)$, where $\bar{\mathbf{F}}=\frac{1}{T} \sum_{t=1}^T \mathbf{F}_t$, and $\mathbf{\Sigma}_F(h)$ is a finite positive definite matrix.

\end{enumerate}

\item Idiosyncratic components

\begin{enumerate}
\item For all $i,j \in \mathbb{N}$ and all $t,h \in \mathbb{Z}$, $\boldsymbol{\varepsilon}_t$ is stationary with zero mean and finite positive-definite symmetric covariance matrix $\mathbf{\Sigma}_{\varepsilon}(0)$ with elements given by $\sigma_{ij}(0):=E\left(\varepsilon_{it} \varepsilon_{jt} \right)$ and auto-cross-covariance matrix for lag $h$, $\mathbf{\Sigma}_{\varepsilon}(h)$ with elements given by $\sigma_{ij}(h):=E\left(\varepsilon_{it} \varepsilon_{jt-h} \right)$. Denote by $\mathbf{\Sigma}_{\varepsilon}:=\mathbf{\Sigma}_{\varepsilon}(0)$, $\sigma_{ij}:=\sigma_{ij}(0)$ and $\sigma^2_i :=\sigma_{ii}(0)$. $| \sigma_{ij}(h) | \le \rho^{|h|} M_{ij}$, where $\rho$ and $M_{ij}$ are finite positive reals independent of $t$, such that $0 \le \rho <1$ and $M_{ii} < C$, $\sum_{j=1,j\ne i}^{N} M_{ij} \le M$ and $\sum_{i=1\ne j}^{N} M_{ij} \le M$, for some finite positive reals $C$ and $M$ independent of $i, j$ and $N$.

\item For all $i$ and $t$, $E[\varepsilon_{it}^8]\leq Q_{\varepsilon}$ for some finite positive real $Q_{\varepsilon}$ independent of $i$ and $t$ independent of $i$ and $t$.

\item For all $j=1,...,N$, all $h=0, 1, 2,...$ and all $T$ and $N$, \\
$E\left[ \left(  \frac{1}{\sqrt{NT}} \sum_{i=1}^N \sum_{t=1}^T \left[  \varepsilon_{it} \varepsilon_{jt} - E(\varepsilon_{it} \varepsilon_{jt}) \right] \right)  ^2 \right] < K_{\varepsilon}$, \\
for some positive real $K_{\varepsilon}$ independent of $j, h, N$ and $T$.

\item The idiosyncratic components and the factors are uncorrelated for all leads and lags.

\end{enumerate}

\end{enumerate}

According to Assumption 1(a), the matrix of loadings is non-random having asymptotically maximum column rank, $r$, and, for any given $N$, all the factor has a finite contribution to each observed series (upper bound on $\parallel \boldsymbol{\lambda}_i \parallel$) and are cross-sectionally pervasive (full rank of $\mathbf{\Sigma}_{\Lambda}$). Under assumptions 1(d) and 1(e), the factors are assumed to be fixed; see Bai and Li (2012, 2016), Onatski (2012), Freyaldenhoven (2022) and Mao \textit{et al}. (2024) for factors treated as fixed unknown constants. If the factors were considered as random, these assumptions imply that the results are conditional on a given realization of a stationary stochastic process; see, for example, the discussion by Barigozzi, Fresoli and Ruiz (2026). Furthermore, the factors are assumed to be pervasive. The idiosyncratic components, $\boldsymbol{\varepsilon}_t$, have finite eight-order moments with full rank covariance matrix, $\mathbf{\Sigma}_{\varepsilon}$, assumed to be positive definite not only for every finite cross-sectional dimension $N$ but also in the limit when $N\rightarrow \infty$. Thus $\mathbf{\Sigma}_{\varepsilon}^{-1}$ exits even in the limit. Existence of the eight-order moment is assumed for the estimator of the covariance matrix of PC loadings to be consistent. Furthermore, note that the idiosyncratic components are allowed to have weak correlations and heteroscedasticity, in both the cross-section and time series dimensions. Note that the assumption of stationarity implies that the unconditional variance of $\varepsilon_{it}$ is constant over time, allowing for conditional heteroscedasticity. Assumption 2(c) is crucial to prove asymptotic normality of the estimated loadings. It also implies that the (auto) covariances between $\varepsilon_{it}$ and $\varepsilon_{jt-h}$ are $\sqrt{T}$-consistent estimators of their population counterparts. 

It is well known that the DFM in (\ref{eq:DFM}) is not identified given that, for any $r \times r$ full rank matrix $\mathbf{H}$, we can define a new set of common factors $\mathbf{F}_t^*=\mathbf{H F}_t $ and loading matrix $\mathbf{\Lambda}^*=\mathbf{\Lambda H}^{-1}$, which can generate the same observed variables in $\mathbf{Y}_t$ as follows
\begin{equation}
\label{eq:identification}
\mathbf{Y}_t=\mathbf{\Lambda H}^{-1} \mathbf{H F}_t + \boldsymbol{\varepsilon}_t = \mathbf{\Lambda}^* \mathbf{F}_t^*+\boldsymbol{\varepsilon}_t.
\end{equation}
Therefore, $r^2$ restrictions are needed in order to identify the loadings and the common factors before proceeding to their estimation. In the context of PC factor extraction, it is common to assume that $\frac{\mathbf{F}'\mathbf{F}}{T}=\mathbf{I}_r$, where $\mathbf{I}_r$ is the $r \times r$ identity matrix, which amounts to $\frac{r(r+1)}{2}$ restrictions. The additional $\frac{r(r-1)}{2}$ identification restrictions are imposed by assuming that $\mathbf{\Lambda}^{\prime} \mathbf{\Lambda}$ is diagonal (with distinct elements in the main diagonal arranged in decreasing order). Under these restrictions, the estimated factors identify the true factors up to a sign transformation. As a consequence, the PC estimated factors and loadings can be directly compared with the corresponding true factors and loadings and there is not need to add a rotation matrix when looking at their asymptotic distributions; see the discussion by Bai and Ng (2013) on factor identification. It is important to remark that, if the identifying restrictions are satisfied, then the estimated PC factors are consistent for the true factors, while, if the restrictions are not satisfied, the estimated factors consistently estimate the space spanned by the true factors and can be rotated to give interpretations of interest. In this latter case, the loadings should also be accordingly rotated. Furthermore, if the factors (loadings) do not satisfy the identification restriction, then the asymptotic distribution is given for $\sqrt{T}\left( \widetilde{\boldsymbol{\lambda}}_{i}-\boldsymbol{\lambda}_{i} \mathbf{H}^{-1}\right)^{\prime}$. Bai and Ng (2013) show that, if the identifying restrictions are satisfied, the rotation matrix is asymptotically an identity matrix.% It is important to note that in case of zeros in the loading matrix, Wei and Zhang (2026) show that the sparsity of the factors is preserve under rotation

Using the identification restrictions described above, the estimated PC factors, $\widetilde{\mathbf{F}}$, are $\sqrt{T}$ times the eigenvectors corresponding to the $r$ largest eigenvalues of $\mathbf{YY}^{\prime}$ arranged in decreasing order. Using the normalization $\frac{\widetilde{\mathbf{F}}^{\prime} \widetilde{\mathbf{F}}}{T}=\mathbf{I}_r$, which is satisfied by construction by PC factors, the PC loadings can be estimated separately for each variable in the system by the following time series regression with the $i$'th cross-sectional unit
\begin{equation}
\label{eq:est-loadings_1}
\widetilde{\boldsymbol{\lambda}}_{i}^{\prime}=\frac{1}{T}\widetilde{\mathbf{F}}^{\prime}\mathbf{Y}_{i \cdot},
\end{equation}
where, for $i=1,...,N$, $\widetilde{\boldsymbol{\lambda}}_{i}$ is the PC estimator of $\boldsymbol{\lambda}_{i}$, the $i$`th row of $\mathbf{\Lambda}$ and $\mathbf{Y}_{i \cdot}$ is the $i$'th column of $\mathbf{Y}$; see Bai (2003) and Bai and Ng (2008).

\subsection{Asymptotic distribution of PC loadings}

Under Assumptions 1 and 2, Bai (2003) establishes consistency of $\widetilde{\boldsymbol{\lambda}}_{i}$, and derives its asymptotic distribution when $N,T\rightarrow \infty$ with $\frac{\sqrt{T}}{N}\rightarrow 0$, which, if the identifying restrictions are satisfied, is given by:
\begin{equation}
\label{eq:asymptotic}
\sqrt{T}\left( \widetilde{\boldsymbol{\lambda}}^{\prime}_{i}-\boldsymbol{\lambda}_{i}^{\prime}\right)\xrightarrow[]{d} N\left( \mathbf{0}, \mathbf{\Omega}_i\right),
\end{equation}
where $\mathbf{\Omega}_i$ is the covariance matrix of the limiting distribution of $\sqrt{T} \left( \frac{1}{T} \sum_{t=1}^T \mathbf{F}_t \varepsilon_{it} \right)$, which is given by%\footnote{Very recently, Wei and Zhang (2026) show that the asymptotic distribution of PC loadings (and factors) do not change when the factors are weak.}
\begin{equation}
\label{eq:Omega}
\begin{split}
\boldsymbol{\Omega}_i &= \lim_{T \rightarrow \infty} \text{Var} \left( \frac{1}{\sqrt{T}} \sum_{t=1}^T \mathbf{F}_t \varepsilon_{it}\right) = \lim_{T \rightarrow \infty}  \frac{1}{T} \sum_{s=1}^T  \sum_{t=1}^T \mathbf{F}_t \mathbf{F}_s^{\prime} E \left(  \varepsilon_{it} \varepsilon_{is}\right) = \sum_{h=-\infty}^{\infty} \boldsymbol{\Sigma}_F(h) \gamma_{\varepsilon_i} (h)\\
&=\mathbf{\Sigma}_F(0) \sigma^2_{i} +  \sum_{h=1}^{\infty} \left[  \mathbf{\Sigma}_F(h) + \mathbf{\Sigma}_F(h)^{\prime}  \right] \sigma_{ii}(h),
\end{split}
\end{equation}
where $\boldsymbol{\Sigma}_F(h)$ is defined in Assumption 1(e) and $\sigma_{ii}(h)$ and $\sigma^2_i$ are defined in Assumption 2(a). If the idiosyncratic components are serially uncorrelated, and the identifying restrictions are satisfied, $\mathbf{\Sigma}_F(0)=I_r$, and consequently,
\begin{equation}
\label{eq:Omega_2}
\mathbf{\Omega}_i=\sigma^2_{i}\mathbf{I}_r
\end{equation}
where $\sigma^2_{i}$ is the marginal variance of the idiosyncratic component of unit $i$, i.e. the $i$'th element of the main diagonal of the idiosyncratic covariance matrix, $\mathbf{\Sigma}_{\varepsilon}$.

As an illustration, consider that as often assumed in the related literature, the serial dependence of the idiosyncratic components is represented by the following stationary AR(1) model
\begin{equation}
\label{eq:idiosy}
\varepsilon_{it}=\delta \varepsilon_{it-1}+ \xi_{it},
\end{equation}
where $|\delta|<1$ and $\xi_{it}$ is a white noise with variance $\sigma^2_{\xi i}$. Then, the covariances of the idiosyncratic components are given by
\begin{equation}
\label{eq:covidiosy}
{\sigma}_{ii}(h)= \delta^h \frac{\sigma^2_{\xi i}}{1- \delta^2}.
\end{equation}
Note that if the idiosyncratic components are further GARCH(1,1), with $\xi_{it}=\nu_{it} \sigma^2_{it}$, where $\nu_{it}$ is white noise with variance equal to unity and $\sigma^2_{it}= \omega_i + \alpha \xi^2_{it-1} + \beta \sigma^2_{it-1}$, then the autocorrelation function of the idiosyncratic components does not change and their variances are modified to be $\sigma^2_{i}=\frac{\omega_i}{(1-\delta^2)(1-\alpha-\beta)}$. Furthermore, consider that the factors are a realization of the following stationary VAR(1) process,
\begin{equation}
\label{eq:VAR_1}
\mathbf{F}_t = \mathbf{\Phi F}_{t-1}+ \boldsymbol{\eta}_t,
\end{equation}
where $\mathbf{\Phi}$ satisfies the stationarity conditions and $\boldsymbol{\eta}_t$ is an $r \times 1$ vector white noise with covariance matrix  $\boldsymbol{\Sigma}_{\eta}$. In this case,
\begin{equation}
\label{eq:covarianza}
\mathbf{\Gamma}_F(h) = \boldsymbol{\Phi}^h \mathbf{\Gamma}_F(0),
\end{equation}
where $\mathbf{\Gamma}_F(0)$ is such that $\text{vec}(\mathbf{\Gamma}_F(0)) = \left( I_{r^2} - \mathbf{\Phi} \bigotimes \mathbf{\Phi}\right)^{-1} \text{vec} (\mathbf{\Sigma}_{\eta})$. When $r=1$ and the identification restrictions are satisfied, the (auto)covariance matrix in (\ref{eq:covarianza}) reduces to $\mathbf{\Gamma}_F(h)= \phi^h$, and the asymptotic variance of the loading of the $i$'th unit is given by
\begin{equation}
\label{eq:covariance_1}
\Omega_i= \sigma^2_i \frac{1 - \phi^2 \delta^2}{(1-\phi \delta)^2} .
\end{equation}

Consider that $\delta=0$ is wrongly assumed when computing the variance of the $i$'th loading, which is calculated as $\Omega^*_i=\sigma^2_i$, then
\begin{equation}
\label{eq:miss}
\Omega^{*}_i-\Omega_i = \frac{2 \sigma_i^2 \phi \delta}{\phi \delta - 1}.
\end{equation}

The sign of the difference between the wrong and the true variances of the loading in (\ref{eq:miss}) can be positive of negative depending on the values of $\phi$ and $\delta$. However, note that the denominator of (\ref{eq:miss}) is always negative. Given that in empirical applications, the signs of $\delta$ and $\phi$ are both expected to be positive, the variance of the loading is usually underestimated when wrongly assuming that the idiosyncratic components are white noise. Consider, for example, that $\phi=0.7$ and $\delta=0.5$, then $\Omega^{*}_i-\Omega_i =-1.077 \sigma^2_i$, with the underestimation of the variance of the loading being larger, the larger is the idiosyncratic variance. 

\section{Feasible testing and confidence intervals for loadings}
\label{sec:testing}

In this section, we describe the available estimators of the covariance matrix of PC loadings available in the literature and propose a new estimator with better finite sample properties.

\subsection{Estimators of the asymptotic covariance matrix of PC loadings}

Note that the asymptotic distribution of the loadings in (\ref{eq:asymptotic}) could be used for inference of the loadings if the parameters governing the idiosyncratic components and the factors were known. However, in practice, these parameters are unknown and feasible inference relies on estimated covariance matrices of PC loadings. In this section, we describe the estimators proposed by Bai (2003), which are often implemented in the related literature, and propose extensions with better finite sample properties.

Assuming that the idiosyncratic components are serially uncorrelated, the covariance matrix of PC loadings can be estimated as follows
\begin{equation}
\label{eq:var}
\widehat{\mathbf{\Omega}}_i^{(0)}=\hat{\sigma}^2_{i}\mathbf{I}_r=\frac{1}{T} \sum_{t=1}^T \widetilde{\varepsilon}_{it}^2 \mathbf{I}_r,
\end{equation} 
where $\widetilde{\varepsilon}_{it}=Y_{it}-\widetilde{\boldsymbol{\lambda}}_{i}\widetilde{\mathbf{F}}_{t}$ are the idiosyncratic residuals. Alternatively, if the idiosyncratic components are serially correlated, Bai (2003) proposes estimating $\mathbf{\Omega}_i$ using the idiosyncratic heteroscedasticity and autocorrelation robust HAC nonparametric estimator proposed by Newey and West (1987) in the context of regression models with serially autocorrelated noises; see Lin and Shin (2023) for an application. The estimator is based on the Bartlett (also known as triangular) kernel as follows:
\begin{equation}
\label{eq:Phi_Bartlett}
\widehat{\mathbf{\Omega}}_i^{(B)}= \hat{\sigma}^2_i \mathbf{I}_r+ \sum_{h=1}^q \left( 1-\frac{h}{q+1}\right) \mathbf{D}_{hi} \mathbf{D}_{hi}^{\prime},
\end{equation}
where $\mathbf{D}_{hi}=\frac{1}{T} \sum_{t=h+1}^T \widetilde{\varepsilon}_{it} \widetilde{\varepsilon}_{it-h} \widetilde{\mathbf{F}}_t \widetilde{\mathbf{F}}_{t-h}^{\prime}$, and $q$ goes to infinity as $T$ goes to infinity. Bai (2003) proves the consistency of $\widehat{\mathbf{\Omega}}^{(B)}_i$ when $qT^{-\frac{1}{4}} \rightarrow 0$.

In the context of regression models, Kolokotrones, Stock and Walker (2024) suggest that the Newey-West estimator based on the Bartlett kernel in (\ref{eq:Phi_Bartlett}) may achieve some optimality among estimators using first-order kernels. In this context, it is well known that the MSE of first-order kernels as the Bartlett kernel is asymptotically dominated by that of second-order kernels, as the quadratic spectral (QS) kernel of Epanechnikov (1969); see, for example, Kolokotrones, Stock and Walker (2024) and the references therein, who show that there can be no optimal first order kernel. Consequently, to improve efficiency in the estimation of $\Omega_i$, we follow Andrews (1991) and propose estimating $\Omega_i$ using the following estimator based on the QS kernel
 \begin{equation}
\label{eq:Phi_QS}
\widehat{\mathbf{\Omega}}_i^{(QS)} = \hat{\sigma}^2_i \mathbf{I}_r+ \sum_{h=1}^q \frac{25(q+1)^2}{12 \pi^2 h^2 A} ( \sin( A) - A \cos (A) ) \mathbf{D}_{hi} \mathbf{D}_{hi}^{\prime},
\end{equation}
where $A=\frac{6 \pi h}{5 (q+1)}$. The QS kernel in (\ref{eq:Phi_QS}) decays more slowly than the Bartlett kernel in (\ref{eq:Phi_Bartlett}).

Regardless of whether the Bartlett kernel in (\ref{eq:Phi_Bartlett}) or the QS kernel in (\ref{eq:Phi_QS}) are used to estimate $\Omega_i$, the thorny issue in practice  is the choice of the lag order $q$; see, for example, the discussion by Volgelsang (2018).  Efficiency in the estimation of $\mathbf{\Omega}_i$ can be improved by choosing optimal rules as those established by Andrews (1991) and Newey and West (1994), which are designed to minimize the MSE of the robust variance estimation the context of regression models and rely on assuming that $\widetilde{\mathbf{F}}_t \widetilde{\varepsilon}_{it}$ follows an AR(1) model. The improvements are larger the stronger the serial correlation of the factors and the idiosyncratic components. Note that an approach that relies on positive serial autocorrelation may not have universal appeal; see, for example, the discussion by West (2018). In this paper, we consider $q=T^{1/5}$. This choice is justified by the fact that it satisfies the conditional for consistency of $\widehat{\mathbf{\Omega}}^{(B)}_i$ established by Bai (2003) and avoids different selections of $q$ for different cross-sectional units, which can complicate the procedure when $N$ is large.

For each cross-sectional unit, $i=1,,,.N$, and factor, $j=1,...,r$, feasible point-wise confidence intervals with coverage $(1-\alpha)\%$ can be constructed for each loading, $\lambda_{ij}$, using the asymptotic normality in (\ref{eq:asymptotic}) with $\mathbf{\Omega}_i$ substituted by $\widehat{\mathbf{\Omega}}_i^{(QS)}$, as follows:
\begin{equation}
\label{eq:confi-bounds}
\widetilde{\lambda}_{ij} \pm \frac{z_{1-0.5 \alpha}}{\sqrt{\hat{\omega}_{i}^{(jj)}}},
\end{equation}
where $z_{\alpha}$ is the $\alpha\%$-quantile of the standard normal distribution and $\hat{\omega}_{i}^{(jj)}$ is the $j$`th element of the main diagonal of $\hat{\mathbf{\Omega}}_i^{(QS)}$. %In particular, assuming that $\lambda_{ij}=0$, one can construct 95\% significance bounds for $\lambda_{ij}$ as follows
%\begin{equation}
%\label{eq:sig_bounds}
%\pm \frac{1.96}{\sqrt{\hat{\omega}_{i}^{(jj)}}}.
%\end{equation}
%The bounds in (\ref{eq:sig_bounds}) are similar to those often used for the sample autocorrelations.
Similarly, one can also test for the individual significance of each loading, $H_0: \lambda_{ij}=0$ using the usual "t"-statistic, $t_{ij}=\frac{\widetilde{\lambda}_{ij}}{\sqrt{\widehat{\omega}_{i}^{(jj)}}}$. For a significance level of $\alpha$, the null is rejected when $t_{ij}>z_{1-0.5\alpha}$.

Finally, one can also obtain joint confidence regions for two or more loadings within $\boldsymbol{\lambda}_{i}$, and test whether they are jointly zero, i.e. $H_0: \mathbf{R} \boldsymbol{\lambda}_{i}^{\prime}=\mathbf{0}$, using the following statistic
\begin{equation}
\label{eq:test1}
T\left(\mathbf{R} \widetilde{\boldsymbol{\lambda}}_{i}^{\prime} \right) \left(\mathbf{R} \widehat{\mathbf{\Omega}}_i^{(QS)} \mathbf{R}^{\prime} \right) ^{-1} \left(\mathbf{R} \widetilde{\boldsymbol{\lambda}}_{i}^{\prime} \right)^{\prime} \xrightarrow[]{d} \chi^2_k,
\end{equation}
where $\mathbf{R}$ is an $1 \times r$ vector of zeros and ones in the position of the loadings that are tested under the null, $k$ is the number of ones and $\chi^2_k$ is the appropriate quantile of the $\chi^2$ distribution with $k$ degrees of freedom.

\subsection{HAR (fixed-b) inference}

The confidence bounds and tests described above are based on the asymptotic normality of PC loadings in (\ref{eq:asymptotic}) with the covariance matrix $\mathbf{\Omega}_i$ substituted by $\widehat{\mathbf{\Omega}}_i^{(QS)}$ (or $\widehat{\mathbf{\Omega}}_i^{(0)}$ if the idiosyncratic components are assumed to be serially uncorrelated), with the bandwidth chosen to guarantee its consistency. However, the sampling bias and variance of $\widehat{\mathbf{\Omega}}^{(QS)}_i$ does not appear in the covariance of $\widetilde{\boldsymbol{\lambda}}_{i}$. As a consequence, in finite samples, unless the temporal sample size, $T$, is very large, these tests tend to incorrectly reject the null too often (the confidence intervals tend to be too tiny), hindering the interpretation of the corresponding factors; see Muller (2014) for a survey in the context of time series regression models with serially correlated errors. This size distortion can be reduced when taking into account the increased variability due to the estimation of $\Omega_i$ by using larger truncation parameter $q$ (Kiefer, Vogelsang and Bunzel, 2000), and/or by using the so called fixed-b critical values of Kiefer and Vogelsang (2005), who propose choosing the bandwidth of the estimator of the covariance matrix as a fixed proportion of the temporal dimension, $q=\lfloor bT \rfloor$, with $b \in (0,1]$. For a given kernel, the choice of truncation parameter trades off the null rejection rate of a test and its power, and this tradeoff differs across kernels. This particular choice leads to a distribution theory that explicitly captures the kernel used and its bandwidth. Inference based on this theory is popularly known as HAR inference; see Jansson (2004), Sun, Phillips and Jin (2008) and Sun (2014), who show that fixed-b critical values provide a higher-order refinement to the null rejection rate of HAR tests in a Gaussian location model.\endnote{Alternatively, one can use Equally Weighted Cosines (EWC), which has standard critical values; see, for example, the comments by Muller (2007, 2018). EWC is closely related to periodogram averaging but cannot be written as a weighted sum of autocovariances.}  Note that the choice of $q$ represents a classic trade-off: small values of $q$ lead to robust tests with low power, while choosing $q$ large leads to powerful tests with potentially large size distortions; see Kiefer and Vogelsang (2005). Sun, Phillips, and Jin (2008) propose bandwidth-selection rules that balance size distortions, particularly over-rejection, against statistical power. Lazarus et al. (2018) and Sun and Yang (2020) develop related procedures in which the bandwidth choice depends on the autocorrelation structure of the data. Following the approach of these latter studies, and in the absence of a bandwidth-selection method specifically calibrated to our settng, in the simulation exercises, we use $b=(0.02, 0.16, 0.20)$ when the autoregressive parameter of the idiosyncratic component is 0, 0.5, and 0.9, respectively, and the Bartlett kernenl is implemented. For the QS kernel, the corresponding values are $b=(0.02, 0.17, 0.22)$.
These values of $b$ may be relatively large, particularly for large temporal dimensions $T$. Nevertheless, as shown later, they yield satisfactory performance of the variance estimators across the designs considered. A dedicated calibration analysis aimed at determining the optimal values of $b$ for our specific setting would be desirable. Such an analysis, however, lies beyond the scope of the present study and is left for future research.%Finally, the 5\% critical values are obtained using the procedure proposed by Sun, Phillips, and Jin (2008), who propose bandwidth rules that balance over-rejections against power, while .
\endnote{Note that Lazarus \textit{et al.} (2018) Lazarus \textit{et al}. (2018) propose a similar rule that depends on the autocorrelation structure of the data. In particular, they  suggest $q=1.3 \sqrt{T}$ assuming that $\widehat{F_t} \widehat{\varepsilon}_{it}$ is an AR(1) model with parameter 0.7 recommend that the null should not be imposed when estimating the long-run variance. Imposing it can result in reductions of size distortions without large losses of power as long as the bandwidth is relatively small and the sample size is not too small. In contrast, the use of large bandwidths while imposing the null can have substantial negative implications for power regardless of the sample size; see Volgelsang (2018).}
Finally, in this paper, the fixed-$b$ critical values are obtained using the procedure proposed by Sun, Phillips and Jin (2008); see also the recommendations by Lazarus \textit{et al}. (2018).

\subsection{Subsampling correction of the asymptotic distribution}
\label{sec:Subsampling}

Another issue when using the (estimated) asymptotic covariance matrix of PC loadings for inference is that it does not take into account that the factors, $F_t$, are unobserved and should be estimated, adding additional uncertainty to the estimated loadings, which is particularly relevant when the cross-sectional dimension, $N$, is not large. Consequently, in this subsection, we propose a subsampling procedure to modify the asymptotic covariance matrix of PC loadings by taking into account that they are estimated using estimated factors instead of the corresponding true unobserved factors.\endnote{Babamoradi, van der Berg and Rinnan (2013) describe the literature using bootstrap procedures in the context of factor models without a temporal dimension.}

The subsampling algorithm proposed in this paper is based on the same arguments as those by Maldonado and Ruiz (2021), who propose correcting the finite sample approximation of the asymptotic covariance matrix of PC factors taking into account the uncertainty associated with the estimation of the factor loadings. Note that, if the true factors, $\mathbf{F}$,  were observed, and the identifying restrictions were satisfied, the loadings could be estimated as follows
\begin{equation}
\label{eq:est-loadings_2}
\widetilde{\boldsymbol{\lambda}}_{i}^{(0)\prime}=\frac{1}{T}\mathbf{F}^{\prime}\mathbf{Y}_{i\cdot}.
\end{equation}
Consequently, the estimation error in the loadings can be decomposed into the error due to using estimated factors, $\widetilde{\mathbf{F}}$, instead of the true factors, $\mathbf{F}$, and the regression estimation error as follows
\begin{equation}
\label{eq:decomposition}
\widetilde{\boldsymbol{\lambda}}_{i}^{\prime}-\boldsymbol{\lambda}_{i}^{\prime} = \left( \widetilde{\boldsymbol{\lambda}}_{i}- \widetilde{\boldsymbol{\lambda}}^{(0)}_{i} \right)^{\prime} + \left( \widetilde{\boldsymbol{\lambda}}^{(0)}_{i} -\boldsymbol{\lambda}_{i}  \right)^{\prime}= \frac{1}{T} \left( \widetilde{\mathbf{F}} - \mathbf{F}\right)^{\prime} \mathbf{Y}_{i \cdot} +\left( \widetilde{\boldsymbol{\lambda}}^{(0)}_{i} -\boldsymbol{\lambda}_{i}  \right)^{\prime} .
\end{equation}
The first component in (\ref{eq:decomposition}) vanishes asymptotically with the cross-sectional dimension, $N$, due to the consistency of PC factors, while the asymptotic distribution in (\ref{eq:asymptotic}) deals with the second component; see Wei and Zhang (2026), who show that estimation errors are uniformly vanishing. However, in finite samples, if $N$ is not very large, then the uncertainty of the PC loadings is also affected by the uncertainty associated to the estimation of the factors.

The proposed correction of the asymptotic covariance matrix of PC loadings is based on subsampling subsets of size $N^*$ of series in the cross-sectional space, with each series containing all temporal observations. For each subsample,  $b=1,...,B$, the factors are estimated by PC, obtaining $\widetilde{\mathbf{F}}^{*(b)}$. For each cross-sectional unit, $i=1,...,N$, the subsampling analogue of the covariance matrix of the loadings due to using estimated factors instead of true factors, is estimated as follows: 
\begin{equation} \label{eq:parameter_2}
\frac{1}{B T^2}\sum_{b=1}^{B}\left( \left( \widetilde{\mathbf{F}}^{\ast (b)}-\widetilde{\mathbf{F}}\right)^{\prime} \mathbf{Y}_{i\cdot} \mathbf{Y}_{i\cdot}^{\prime} \left( \widetilde{\mathbf{F}}^{\ast (b)}-\widetilde{\mathbf{F}}\right) \right).
\end{equation}
Finally, the finite sample covariance matrix of PC loadings is estimated as follows:
\begin{equation} \label{eq:Subsampling_MSE_2}
\widehat{Avar}^{\ast}(\widetilde{\boldsymbol{\lambda}}_{i}) =\frac{1}{T} \widehat{\mathbf{\Omega}}^{(QS)}_i + \frac{1}{B T^2}\sum_{b=1}^{B}\left( \left( \widetilde{\mathbf{F}}^{\ast (b)}-\widetilde{\mathbf{F}}\right)^{\prime} \mathbf{Y}_{i\cdot} \mathbf{Y}_{i\cdot}^{\prime} \left( \widetilde{\mathbf{F}}^{\ast (b)}-\widetilde{\mathbf{F}} \right)\right).
\end{equation}

\section{Monte Carlo experiments}
\label{sec:MC}

In this section, we analyse the finite sample coverages of pointwise confidence intervals for individual PC loadings based on their asymptotic distribution. We also analyse the performance of individual significance tests for PC loadings. We consider designs with several combinations of the cross-sectional and temporal dimensions, different number of factors, different persistence of the factors, and different structures of the idiosyncratic components tailored to represent DFMs often estimated in empirical applications.

\subsection{Monte Carlo design}

Systems of $N=25, 50, 100, ..., 200$ variables with $T=25, 50, 100,...,200$ observations are generated by the DFM in (\ref{eq:DFM}) with $r=1$ and $r=3$ factors, and with the idiosyncratic components, $\boldsymbol{\varepsilon}_t$, generated by a multivariate Gaussian white noise with covariance matrix $\mathbf{\Sigma}_{\varepsilon}$,  whose elements in the main diagonal are generated by $\sigma_i^2= cu_i$, where $u_i \sim U(0.5, 10)$ and $c=0, 0.5, 1$ and 2, measures the noise-to-signal ratio. The off-diagonal elements of $\mathbf{\Sigma}_{\varepsilon}$ are generated according to the following Toeplitz structure:
\begin{equation}
\label{eq:cross}
\sigma_{ij}=\sigma_i \sigma_j \tau^{j-i}, i=1,...,N, j=i+1,...,N,
\end{equation}
with $\tau=-0.95, -0.9, -0.7, -0.5, -0.3, -0.1, 0, 0.1, 0.3, 0.5, 0.7, 0.9$ and 0.95. Note that, provided that $|\tau|<1$, the condition of weak idiosyncratic cross-correlation is satisfied; see Fresoli, Poncela and Ruiz (2024). When $\tau$ is negative, the columns of $\mathbf{\Sigma}_{\varepsilon}$ are permuted to generate realistic structures of the idiosyncratic cross-correlations close to those often observed in empirical applications. Note that when $|\tau|=0.9$ or 0.95, although the weak cross-correlation of the idiosyncratic components is still satisfied, the correlations between the idiosyncratic components can be so strong that a further common factor can be appropriate. The cases more realistic from an empirical point of view are those for which $|\tau| \leq 0.7$. However, we still consider $|\tau|=0.9$ and 0.95 in our simulations for completeness. We also generate the idiosyncratic components with temporal dependence by the following AR(1) models
\begin{equation}
\label{eq:AR}
\varepsilon_{it}=\delta \varepsilon_{it-1} + \left( 1- \delta^2\right)^{1/2} \nu_{it},
\end{equation}
where $\nu_{it}$ is a Gaussian white noise with variance $\sigma^2_i$ defined as above. We consider $\delta=0, 0.5$ and 0.95. Finally, we also consider conditionally heteroscedastic idiosyncratic components generated as in (\ref{eq:AR}) with $\nu_{it}=\nu^*_{it}\sigma_{it}$, where $\nu^*_{it}$ is Gaussian white noise with variance one, and 
\begin{equation}
\label{eq:GARCH}
\sigma^2_{it} = \sigma^2_i (1-\alpha-\beta) + \alpha \nu^2_{it-1} + \beta \sigma^2_{it-1},
\end{equation}
where the GARCH parameters are chosen to resemble those often encountered in empirical applications, namely $\alpha=0.2$ and $\beta=0.75$. Finally, we also consider idiosyncratic components generated by a Student-7 distribution.
 
With respect to loadings and factors, when $r=1$, the former are generated by $\lambda_{i}\sim U(0.5,1.5)$, while the latter is generated by an AR(1) model with the disturbance given by $u_t\sim N(0, 1-\phi^2)$ with $\phi=0.7$, 0.9 and 1.\endnote{Note that, in this paper, factors are assumed to be weakly stationary. However, Bai (2004) shows that, if the factors are $I(1)$ and the idiosyncratic errors are stationary, PC estimation still provides a consistent estimation of the common factors and factor loadings. The convergence rate for PC factors still is $\sqrt{N}$, but the rate for PC loadings has a fast rate of $T$. Thus, more precise estimation is obtained when factors are $I(1)$. This is intuitive because the information noise ratio is high with $I(1)$ factors. Bai (2004) also derives the limiting distribution of PC factors and loadings, which, conditional on a given realization of the factors, can be approximated in finite samples as that of the loadings in the stationary DFM. Consequently, for completeness, we also consider non-stationary factors in our simulation study. In the case of $\phi=1$, the variance of $u_t$ is 1.} Finally, when the number of common factors is $r=3$, the factors are generated by the following VAR(1) model
\begin{equation}
\label{eq:Factors}
\mathbf{F}_{t}=\mathbf{\Phi F}_{t-1}+ \mathbf{u}_{t},
\end{equation}
where $\Phi= diag(0.7, 0.4, 0.4)$ and $\mathbf{u}_{t}$ is an $3\times1$ Gaussian white noise vector process with covariance matrix given by $\mathbf{\Sigma}_{u}= diag (\frac{1}{1-0.7^2}, \frac{1}{1-0.4^2}, \frac{1}{1-0.4^2})$. The loadings of the first factor are generated as before, while the loadings of the second and third factors are generated by a $U(0,1)$ distribution independently of the loadings of the first factor. The simulated factors are standardized to satisfy the usual identifiability restrictions, namely, $\frac{\mathbf{F}^{\prime}\mathbf{F}}{T}=\mathbf{I}_{r}$, while the loadings of the second and third factors are transformed to satisfy $\mathbf{\Lambda}^{\prime}\mathbf{\Lambda}$ being diagonal. In fact, if $\lambda^*_{i2}$ are the $U(0,1)$ loadings generated for the second factor, we obtain $\lambda_{i2}= \lambda^*_{i2}- \left(\boldsymbol{\lambda}_1^{\prime} \boldsymbol{\lambda}_1 \right) ^{-1} \boldsymbol{\lambda_1}^{\prime} \boldsymbol{\lambda}^*_{2} \lambda_{i1}$, where $\boldsymbol{\lambda}_1$ is the $N \times 1$ vector of loadings of the first factor and $\boldsymbol{\lambda}_2^*$ is the $N \times 1$ vector of loadings generated for the second factor. A similar transformation is carried out for $\lambda_{i3}$.

For each design, the number of replications is $R=2000$. Before extracting the factors, the simulated variables are, as usual, centred and standardized. It is important to note that by standardizing, the heteroscedasticity is reduced and the simulated systems can be considered as approximately homoscedastic; see the discussion by Ruiz and Poncela (2022). The factors and their loadings are estimated by PC and point-wise confidence intervals and tests for each loading are constructed using the asymptotic distribution as in (\ref{eq:asymptotic}) with $\mathbf{\Omega}_i$ estimated assuming serially uncorrelated and homoscedastic idiosyncratic components as in (\ref{eq:var}) or with the HAC estimator in (\ref{eq:Phi_QS}). We also analyse the performance of the subsampling correction. The empirical coverage of 95\% point-wise confidence intervals for each loading as well as the corresponding empirical significance level of individual significance tests are calculated through the Monte Carlo simulations.

\subsection{Results for point-wise confidence intervals and tests}

Consider first the results for a DFM with $r=1$ factor generated by an AR(1) model with $\phi=0.7$, and with Gaussian idisyncratic components such that the noise-to-signal being $c=1$.\endnote{Results for $\phi=0.9$ and 1 are qualitatively the same. They are available upon request. If $c=0.5$, the coverages are obviously closer to the 95\% nominal. Furthermore, results for GARCH(1,1), cross-sectionally correlated, or Student-7 idiosyncratic components are available in Appendix A.} To analyse the effect of the structure of the idiosyncratic components on the empirical coverages of point-wise confidence intervals and individual significance tests for the loadings, we consider cross-sectionally uncorrelated white noise idiosyncratic components, i.e. $\tau=0$, and $\delta=0$ as well as idiosyncratic serial dependence with $\delta=0.5$ and 0.95.

\begin{figure}[h!]
\includegraphics[trim=0cm 0cm 0cm 0.85cm, clip, width=0.45\textwidth]{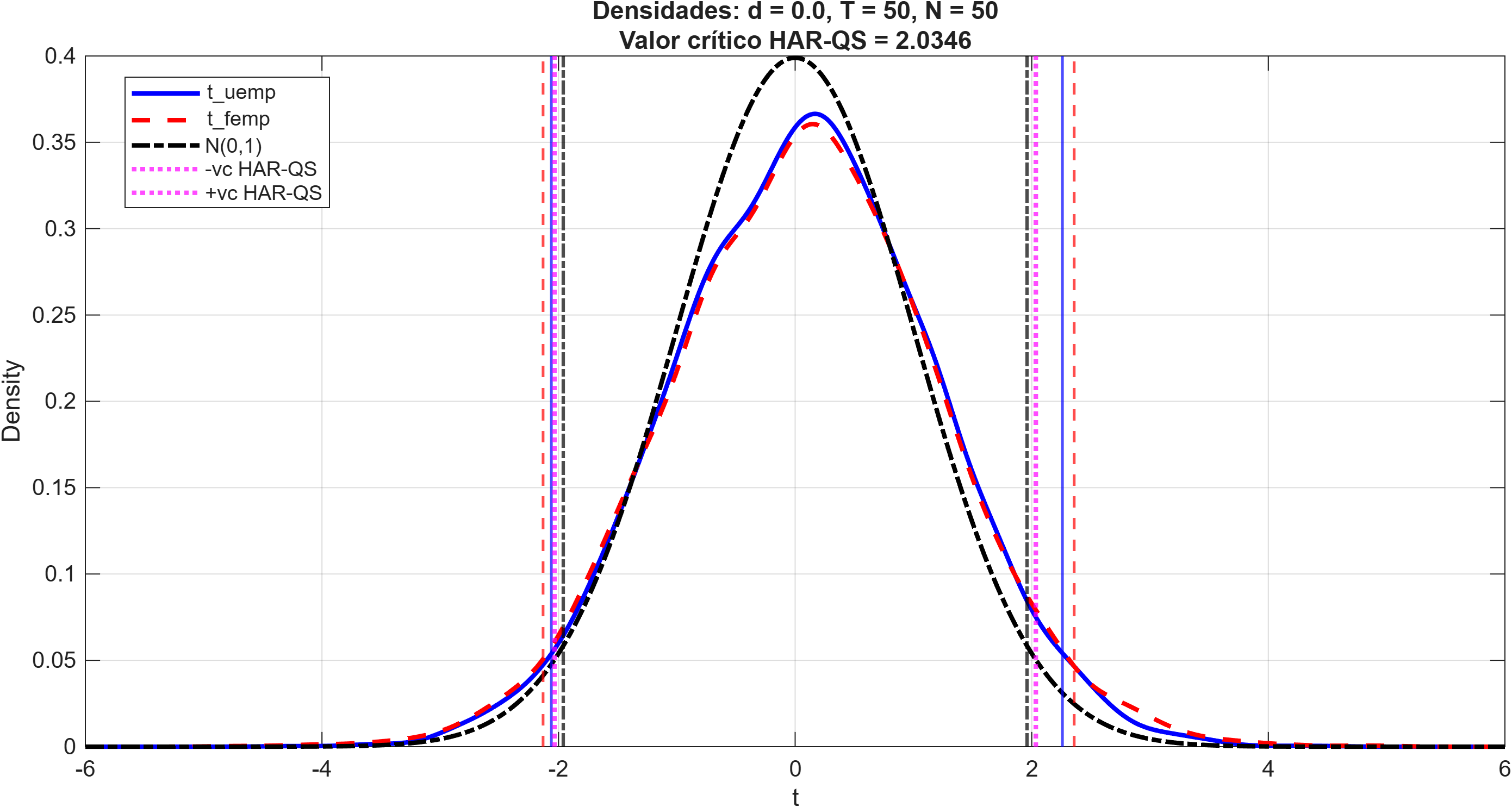}
\includegraphics[trim=0cm 0cm 0cm 0.85cm, clip,width=0.45\textwidth]{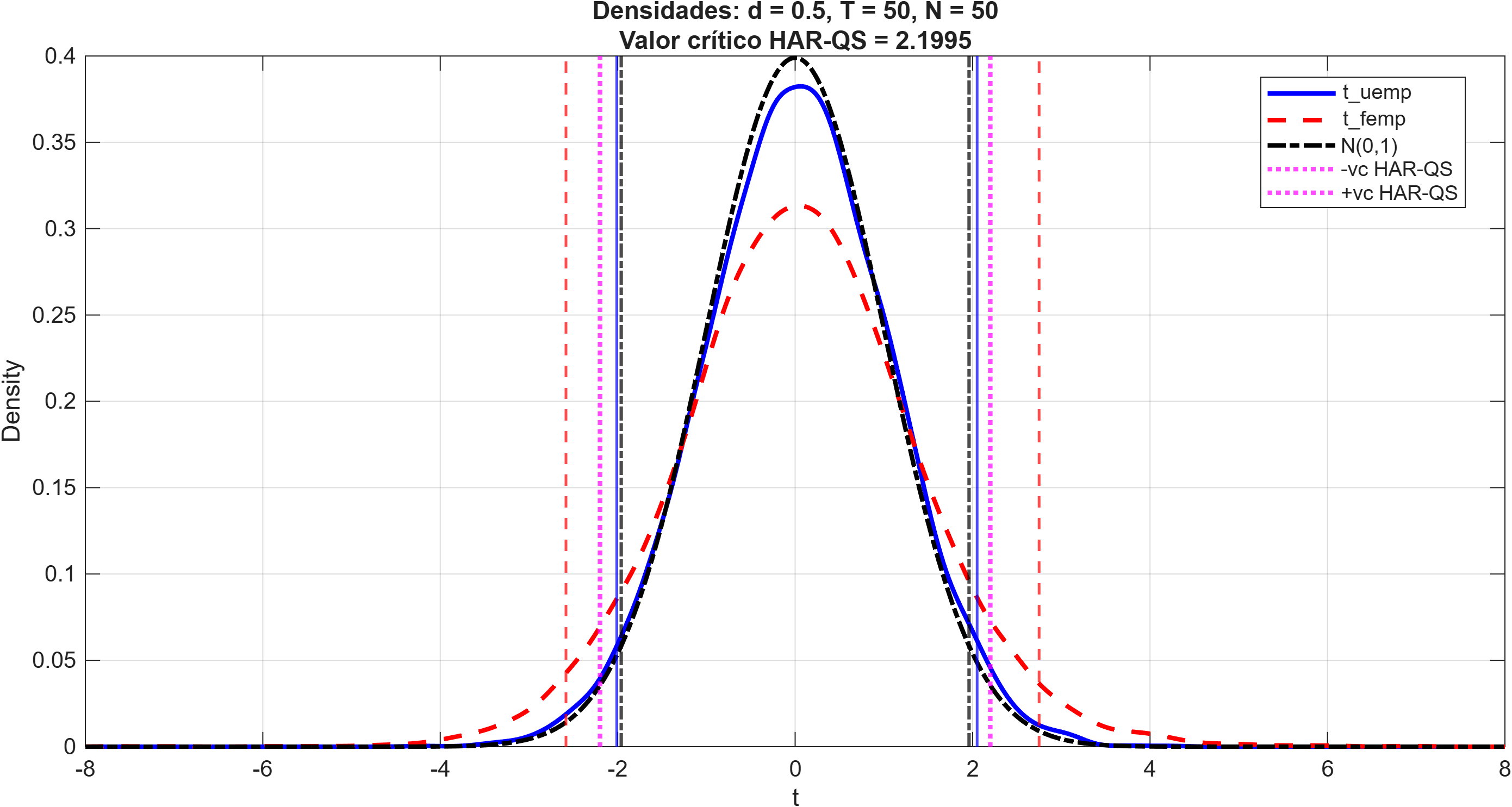}\\
\includegraphics[trim=0cm 0cm 0cm 0.85cm, clip,width=0.45\textwidth]{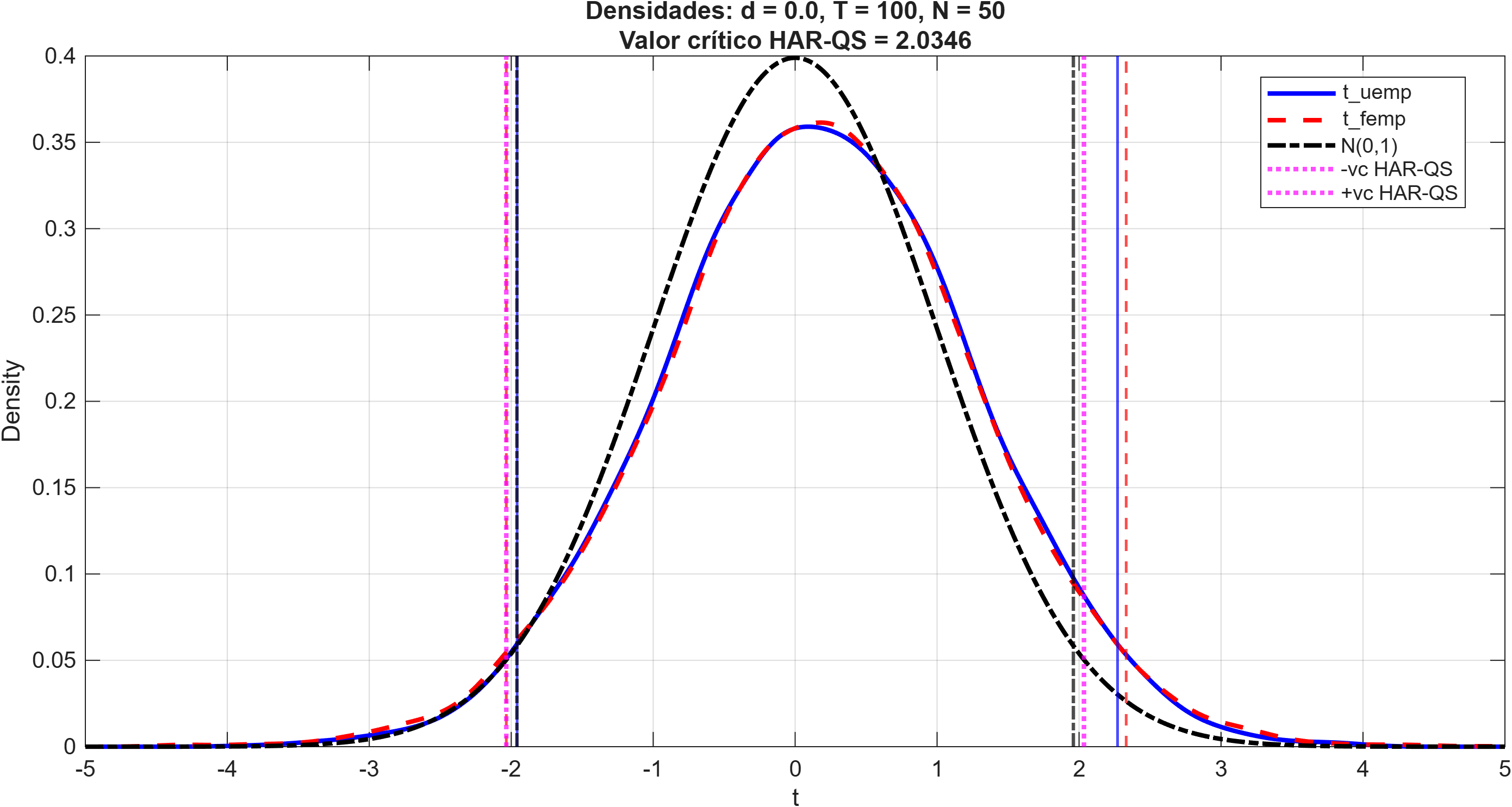}
\includegraphics[trim=0cm 0cm 0cm 0.85cm, clip,width=0.45\textwidth]{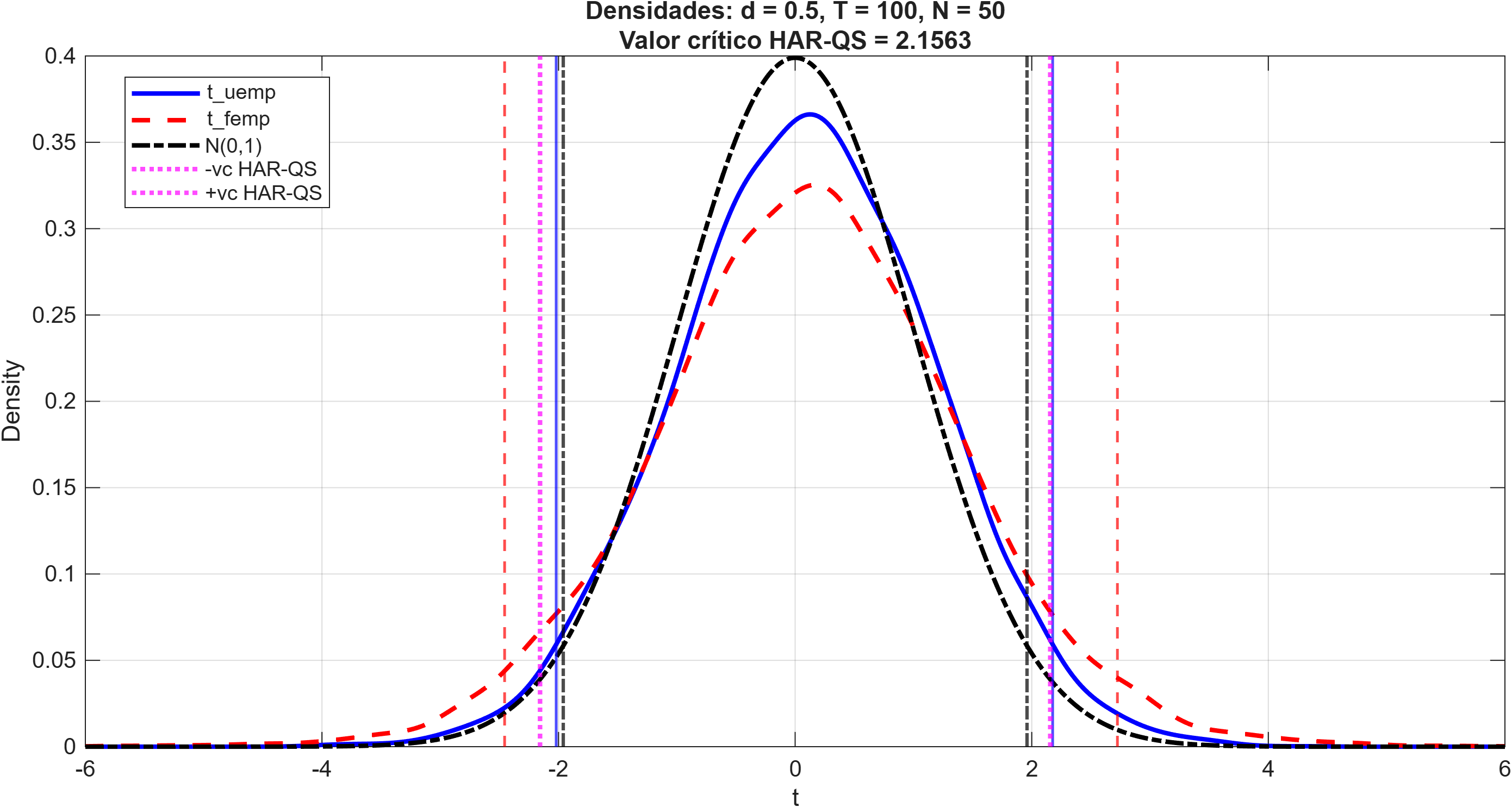}\\
\includegraphics[trim=0cm 0cm 0cm 0.85cm, clip,width=0.45\textwidth]{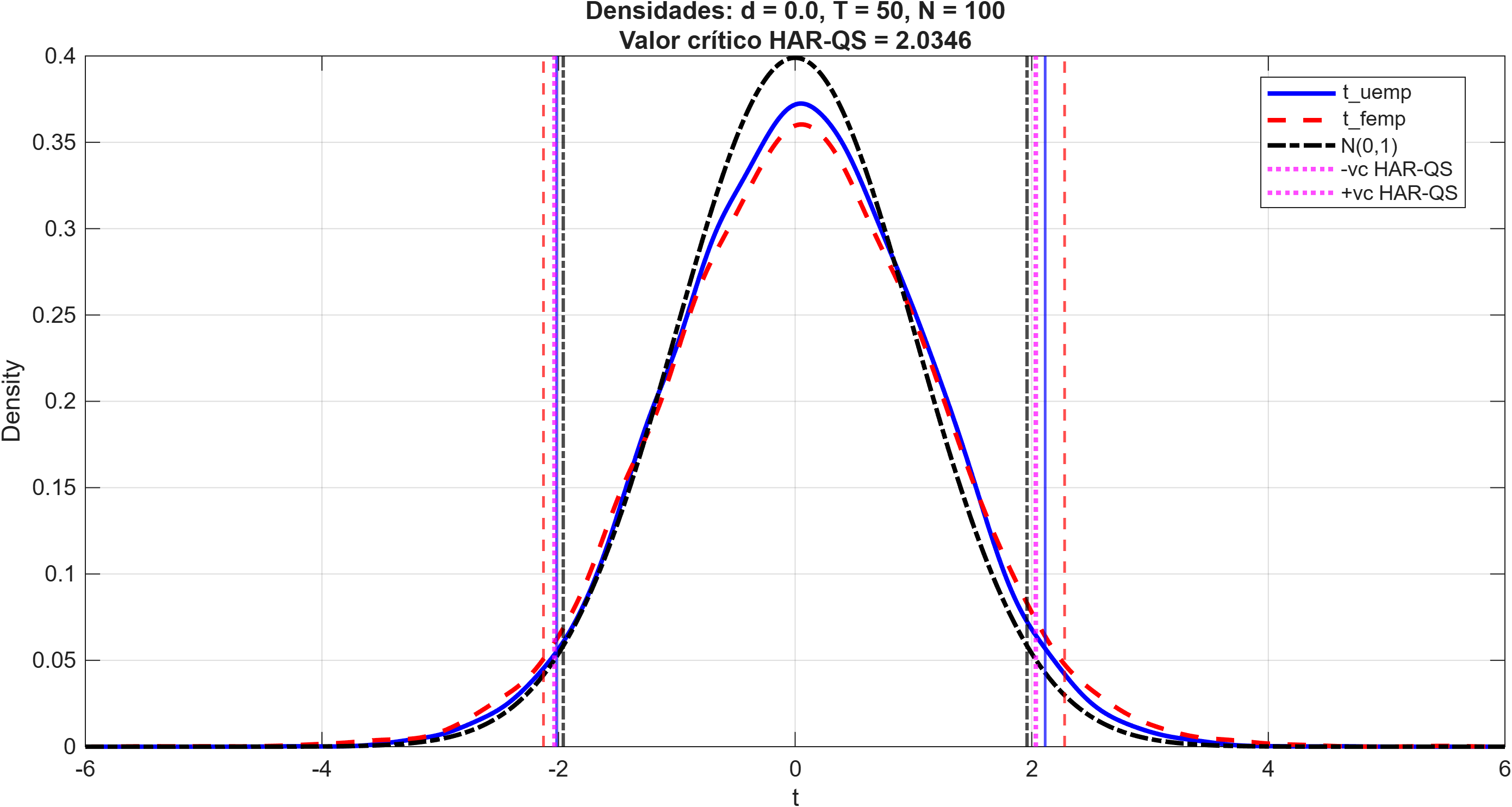}
\includegraphics[trim=0cm 0cm 0cm 0.85cm, clip,width=0.45\textwidth]{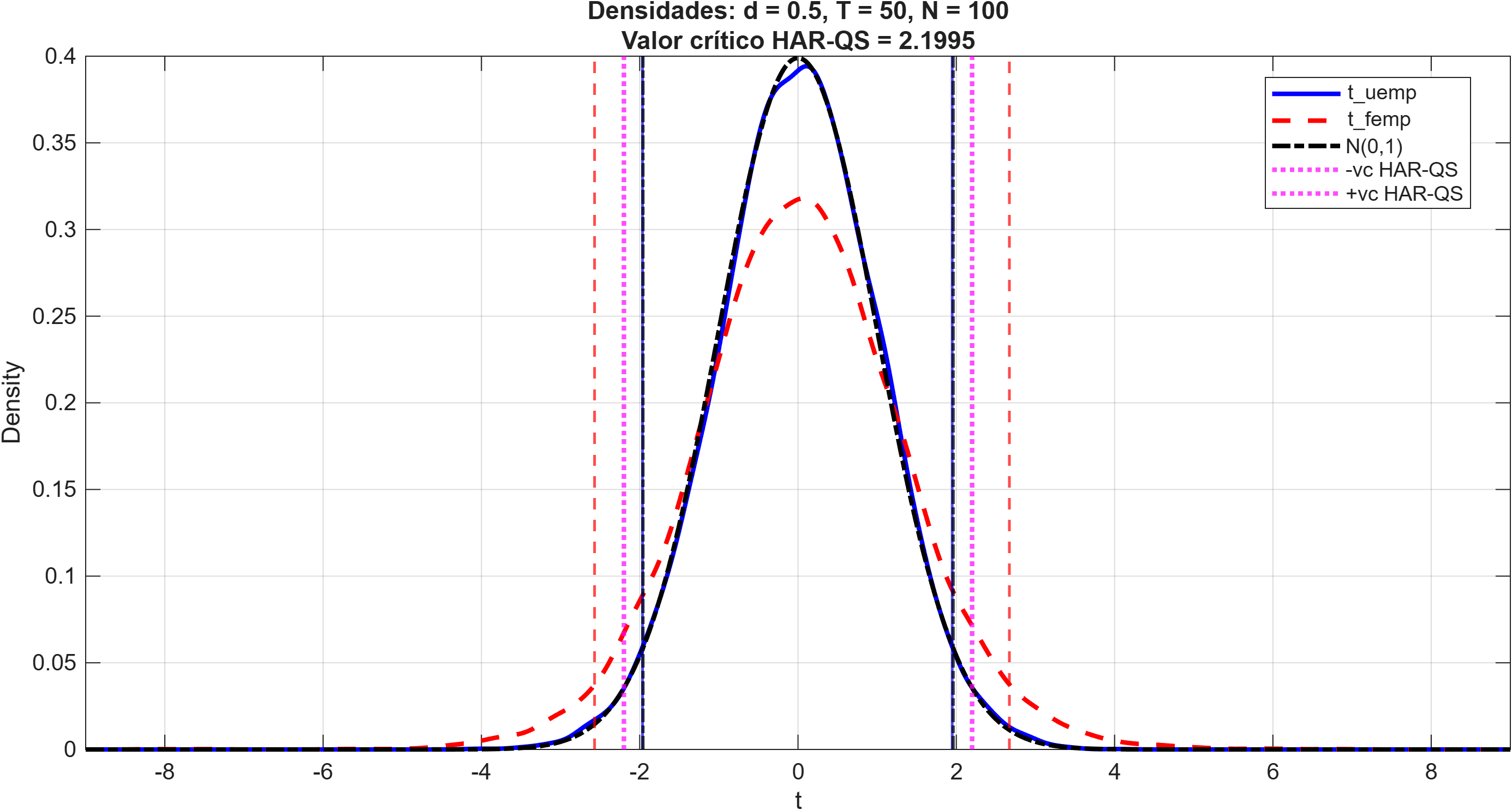}\\
\includegraphics[trim=0cm 0cm 0cm 0.85cm, clip,width=0.45\textwidth]{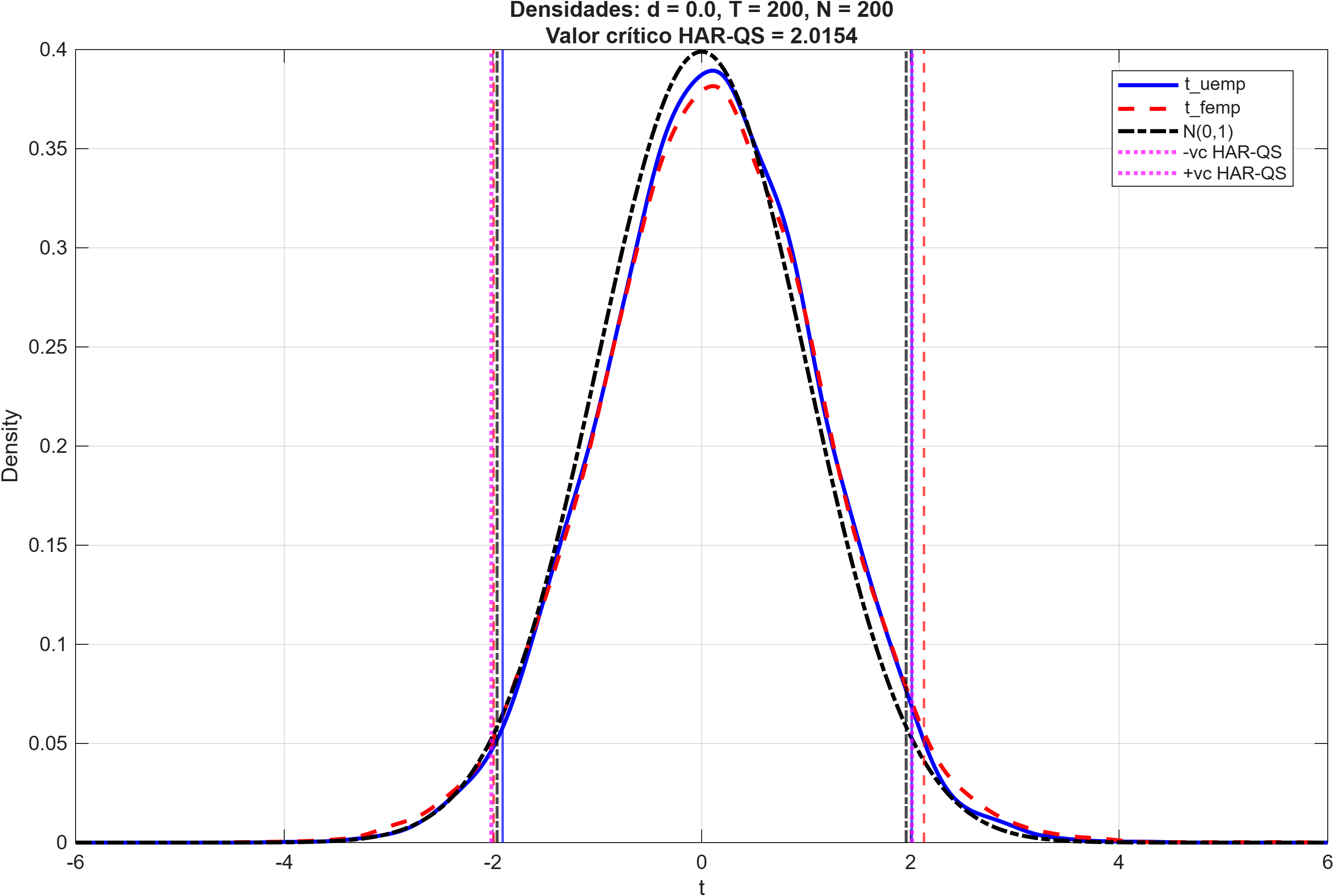}
\includegraphics[trim=0cm 0cm 0cm 0.85cm, clip,width=0.45\textwidth]{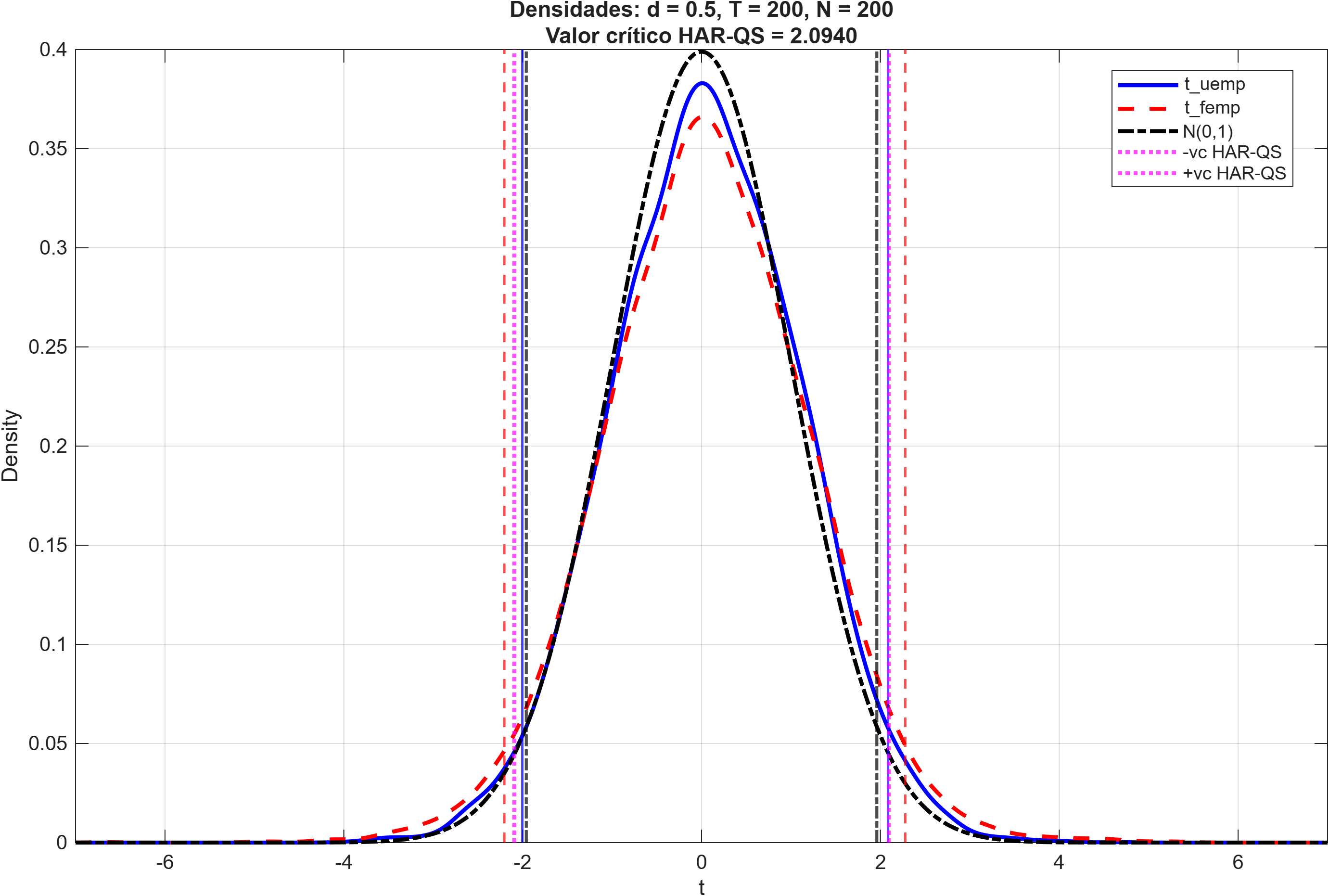}
\vspace{-2mm}
\caption{Empirical densities of standardized PC loading, $\lambda_1$, for the DFM with $r=1$, $\phi=0.7$ and $c=1$, with idiosyncratic components being: cross-sectional and serially uncorrelated (first column); cross-sectionally uncorrelated and serially dependent with $\delta=0.5$ (second column) together with the standard normal density (discontinuous black lines). The sample dimensions are: i) $T=N=50$ (first row); ii) $T=100$ and $N=50$ (second row); $T=50$ and $N=100$ (third row); and iv) $T=N=200$ (fourth row). The asymptotic covariance matrix is computed using known parameters (blue continuous lines) and estimating it using the QS kernel and the subsampling correction (red discontinuous lines). Vertical lines correspond to the 2.5\% and 97.5\% quantiles of each of the densities and HAR quantiles (pink vertical lines).}
\label{fig:densities}
\end{figure}

Figure \ref{fig:densities} plots the empirical Monte Carlo densities of $\widetilde{\lambda}_1$ standardized using $\omega_{1}$ (obtained assuming that $\mathbf{\Omega}_1$ is known), and using $\widehat{\omega}_1$ (estimating $\mathbf{\Omega}_1$ using the QS kernel and the subsampling correction) as defined in (\ref{eq:confi-bounds}). We consider four sets of dimensions: i) small dimensions, $T=N=50$; ii) $T=100$ and $N=50$, which are close to the dimensions in the empirical application below; iii) $T=50$ and $N=100$; and iv) large dimensions, $T=N=200$. We also consider serially uncorrelated ($\delta=0$) and serially correlated  ($\delta=0.5$) idiosyncratic components. We can observe that, regardless of the sample dimensions, the Monte Carlo densities of the PC estimated loadings standardized using the unfeasible (known) variance or using the feasible (estimated) variance are nearly the same when $\delta=0$, i.e. the idiosyncratic components are white noise.  We can also observe that, increasing the cross-sectional dimension $N$ or the temporal dimensions has a very minor effect on the empirical densities.  When $T=N=200$, the empirical densities are very close to the asymptotic standard normal density. When $\delta=0$, we can also observe that, as expected, the standard normal and the HAR quantiles are very close to each other. The picture is similar when the idiosyncratic components are serially correlated and the dimensions are large $T=N=200$. However, when $T=N=50$, the densities obtained using the unfeasible known variance of $\widetilde{\lambda}_1$ or the feasible estimated one, are far apart with the former being very close to the asymptotic standard normal density, while the latter has increased uncertainty. In this case, the HAR quantiles are closer to the empirical quantiles of the distribution of $\frac{\widetilde{\lambda}_1}{\widehat{\omega}_1}$, although they are still smaller in absolute value.   
%while increasing the temporal dimensions $T$, the densities are closer to the asymptotic standard normal distribution. We also show that increasing $N$ or $T$ does not have any effect on this picture, the Monte Carlo empirical distribution of $\frac{\widetilde{\lambda}_1}{\widehat{\omega}_1}$ remains unchanged and far from the asymptotic standard normal distribution. The HAR quantiles are closer to the empirical one, although still smaller in absolute value. Finally, when $T=N=200$, the empirical distribution of $\frac{\widetilde{\lambda}_1}{\widehat{\omega}_1}$ is close to the asymptotic standard normal, with their quantiles being closer to the HAR critical values. 

\begin{figure}[h!]
\includegraphics[ trim=0 0 0 40, clip, width=1.0\textwidth]{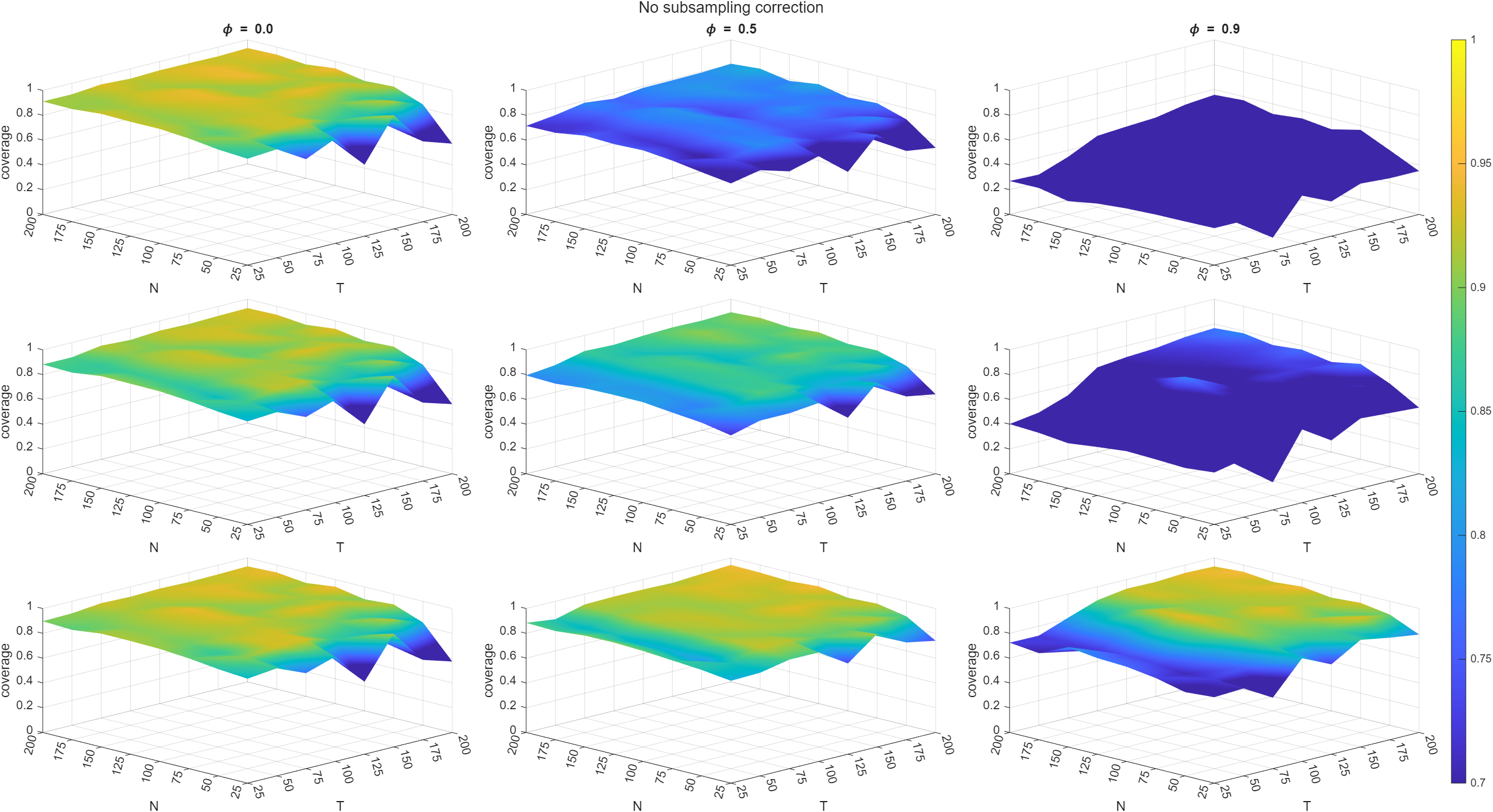}
\vspace{-2mm}
\caption{Empirical coverages of point-wise confidence intervals for $\lambda_1$ for the DFM with $r=1$, $\phi=0.7$ and $c=1$, with idiosyncratic components being: cross-sectional and serially  uncorrelated (first column); cross-sectionally uncorrelated and serially dependent with $\delta=0.5$ (second column); cross-sectionally uncorrelated and serially dependent with $\delta=0.95$ (third column). The asymptotic covariance matrix is computed without (first row) and with the HAC correction using inference based on normality (second row) and on HAR (third row).}
\label{fig:Coverages_ws}
\end{figure}

\begin{figure}[h!]
\includegraphics[trim=0 0 0 40, clip, width=1.0\textwidth]{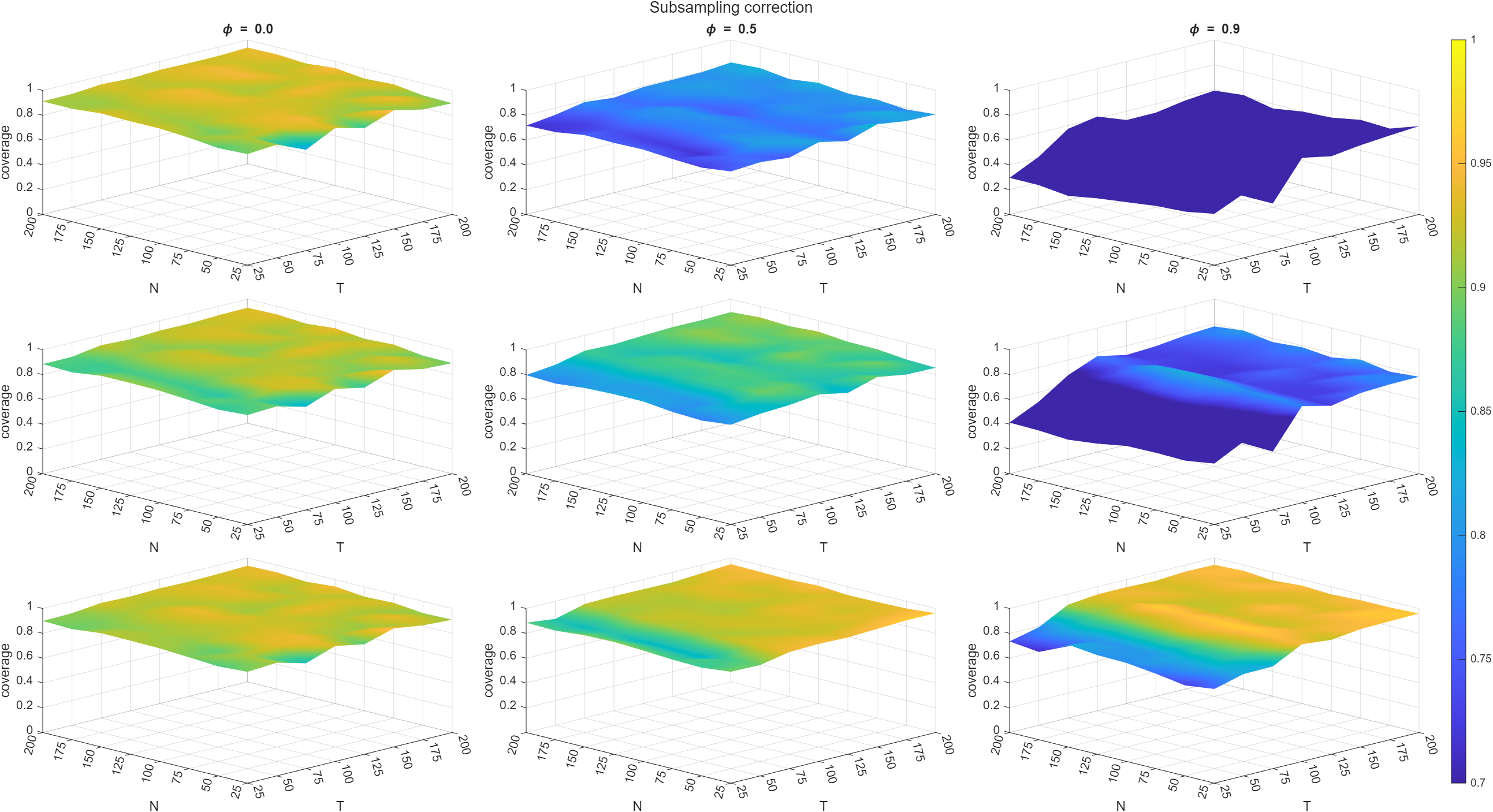}
\vspace{-2mm}
\caption{Empirical coverages of point-wise confidence intervals for $\lambda_1$ for the DFM with $r=1$, $\phi=0.7$ and $c=1$, with idiosyncratic components being: cross-sectional and serially  uncorrelated (first column); cross-sectionally uncorrelated and serially dependent with $\delta=0.5$ (second column); cross-sectionally uncorrelated and serially dependent with $\delta=0.95$ (third column). The asymptotic covariance matrix is computed using the subsampling correction and without (first row) and with the HAC correction using inference based on normality (second row) and on HAR (third row).}
\label{fig:Coverages_s}
\end{figure}

Figures \ref{fig:Coverages_ws} and \ref{fig:Coverages_s} plot the empirical coverages of the point-wise confidence intervals for $\lambda_1$ as a function of the temporal, $T$, and cross-sectional, $N$,  dimensions, when they are constructed based on the covariance matrix of the PC loadings estimated assuming that the idiosyncratic components are white noise ($\widehat{\mathbf{\Omega}}_1^{(0)}$), and with the HAC correction based on the QS kernel and, in the latter case,  with inference based on asymptotic normality and on HAR.\endnote{The Monte Carlo coverages of the intervals constructed using the Bartlett kernel $\widehat{\Omega}^{(B)}_i$ are worse than those obtained using the quadratic kernel $\widehat{\Omega}^{(QS)}_i$ for samples sizes typically encountered in empirical applications of DFMs. They are not reported to save space, but are available upon request.} In Figure \ref{fig:Coverages_ws} there is not subsampling correction, while the coverages plotted in Figure \ref{fig:Coverages_s} are obtained when the estimated covariance matrix is corrected using subsampling. Several empirically relevant conclusions can be extracted from Figure \ref{fig:Coverages_ws}. First, we can observe that the coverages depend strongly on the cross-sectional dimension $N$ and not so much on the temporal dimension $T$. Second, if the idiosyncratic noises are cross-sectionally and serially uncorrelated, the empirical coverages are close to the nominal 95\% regardless of $T$, when $N$ is large. However, for $N=75$, the empirical coverage decreases to approximately 90\%, and, if $N$ is below 50, the coverage can be as low as 80\%. Therefore, a large $N$ is needed for a correct inference on the loadings even if they are estimated by separate regressions for each cross-sectional unit as in (\ref{eq:est-loadings_1}). This is due to the fact that these separate regressions are based on estimated factors, whose precision depend on $N$; see the results on the precision of estimated factors by Poncela and Ruiz (2016), who show that cross-sectional dimensions over $N=30$ are required for the estimated factors to be reasonably precise. This undercoverage is even stronger when there is idiosyncratic cross-correlation, with nominal coverages being around 70\% when $N=25$. Even for large sample sizes, $N=T=200$, the coverage is around 90\%. Once more, the undercoverage is associated to the uncertainty in the estimation of the factors, which is larger when there is idiosyncratic cross-sectional correlation; see Fresoli, Poncela and Ruiz (2023, 2025).

Second, we can observe that, as expected, using the HAC correction improves the coverage in the designs in which the idiosyncratic components are serially autocorrelated, while in those designs in which they are white noise, the empirical coverages are nearly the same regardless of whether the covariance matrix of PC loadings is estimated using (\ref{eq:var}) or the fixed-b critical values. Therefore, in concordance with the results in Figure \ref{fig:densities} representing the Monte Carlo densities, the HAC correction is relevant when computing the asymptotic covariance matrix of the loadings in the presence of idiosyncratic temporal dependence, while it is not worse if they are white noise. Consequently, it is recommended to use it always when computing the asymptotic covariance matrix of PC loadings.

Third, note that, even if the HAC estimator of the covariance matrix is used, the empirical undercoverage of the confidence intervals for the loadings based on using the asymptotic distribution can be severe when the idiosyncratic components are serially correlated. This undercoverage gets worse as the strength of the idiosyncratic autocorrelation increases.  It is also remarkable that only when the serial correlation of the idiosyncratic components is strong, the empirical coverage depends on the temporal size $T$.

\begin{figure}[h!]
\includegraphics[trim=0 0 0 62, clip, width=1.0\textwidth]{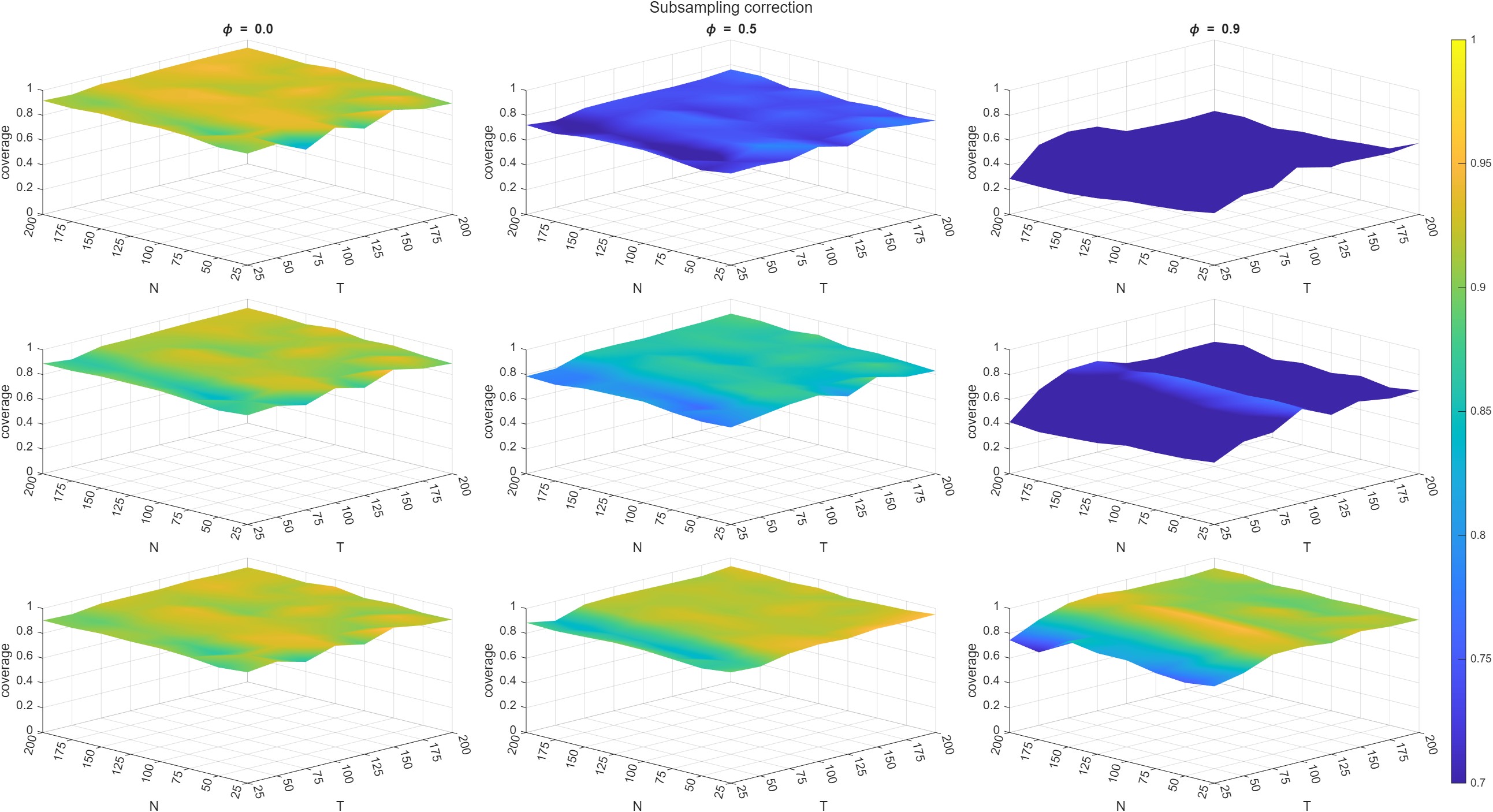}
\vspace{-2mm}
\caption{Empirical coverages of point-wise confidence intervals for $\lambda_1$ for the DFM with $r=1$, $\phi=0.9$ and $c=1$, with idiosyncratic components being cross-sectionally uncorrelated: serially uncorrelated (first column); serially dependent with $\delta=0.5$ (second column); and serially dependent with $\delta=0.9$ (third column). The asymptotic covariance matrix is computed with subsampling correction, without (first row) and with the HAC correction with inference based on normality (second row) and on HAR (third row).}
\label{fig:Coverage_persistent_2}
\end{figure}

\begin{figure}[h!]
\includegraphics[trim=0 0 0 62, clip, width=1.0\textwidth]{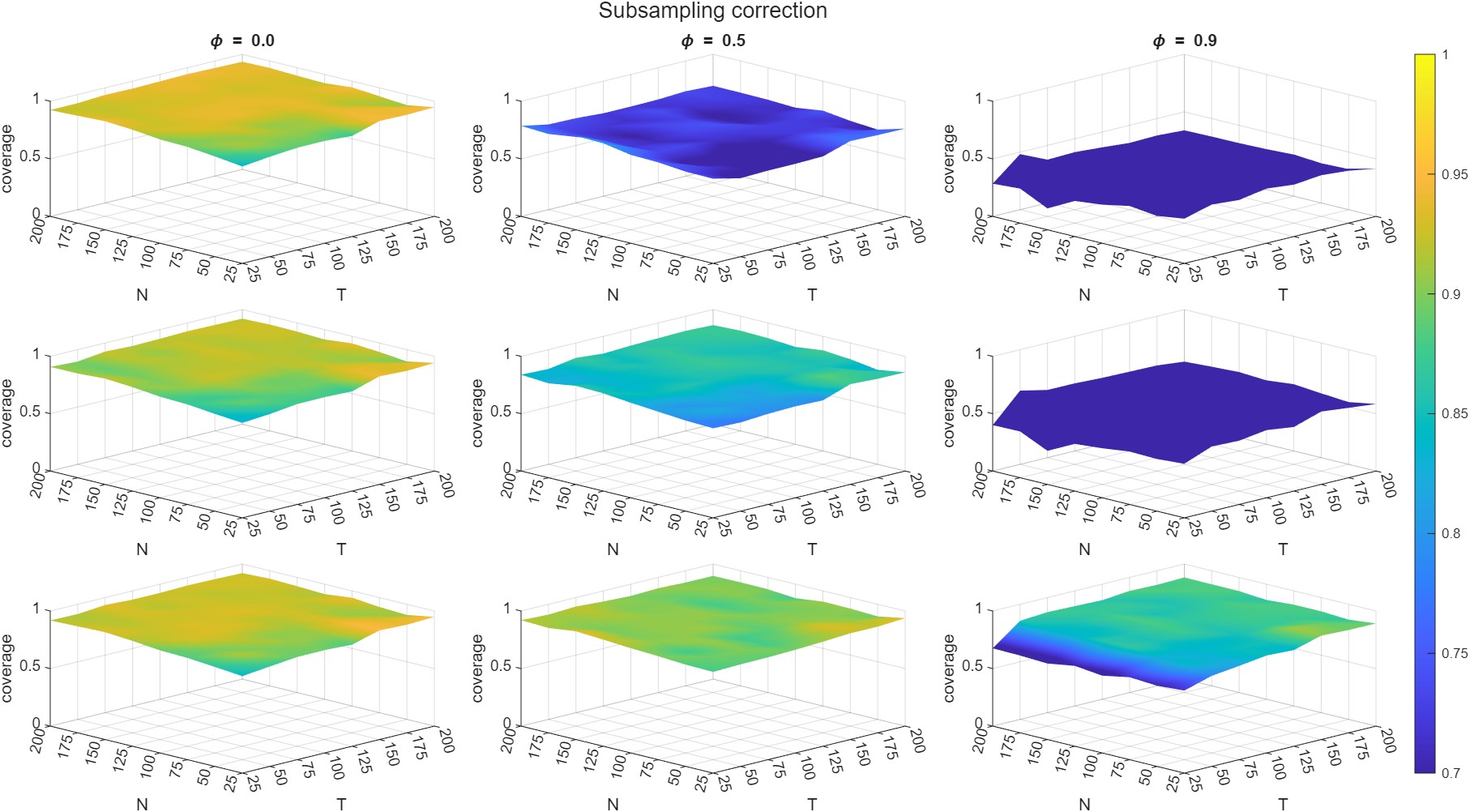}
\vspace{-2mm}
\caption{Empirical coverages of point-wise confidence intervals for $\lambda_1$ for the DFM with $r=1$, $\phi=1$ and $c=1$, with idiosyncratic components being cross-sectionally uncorrelated: serially uncorrelated (first column); serially dependent with $\delta=0.5$ (second column); and serially dependent with $\delta=0.9$ (third column). The asymptotic covariance matrix is computed with subsampling correction, without (first row) and with the HAC correction with inference based on normality (second row) and on HAR (third row).}
\label{fig:Coverage_persistent_3}
\end{figure}

\begin{figure}[h!]
\includegraphics[trim=0 0 0 80, clip, width=0.55\textwidth]{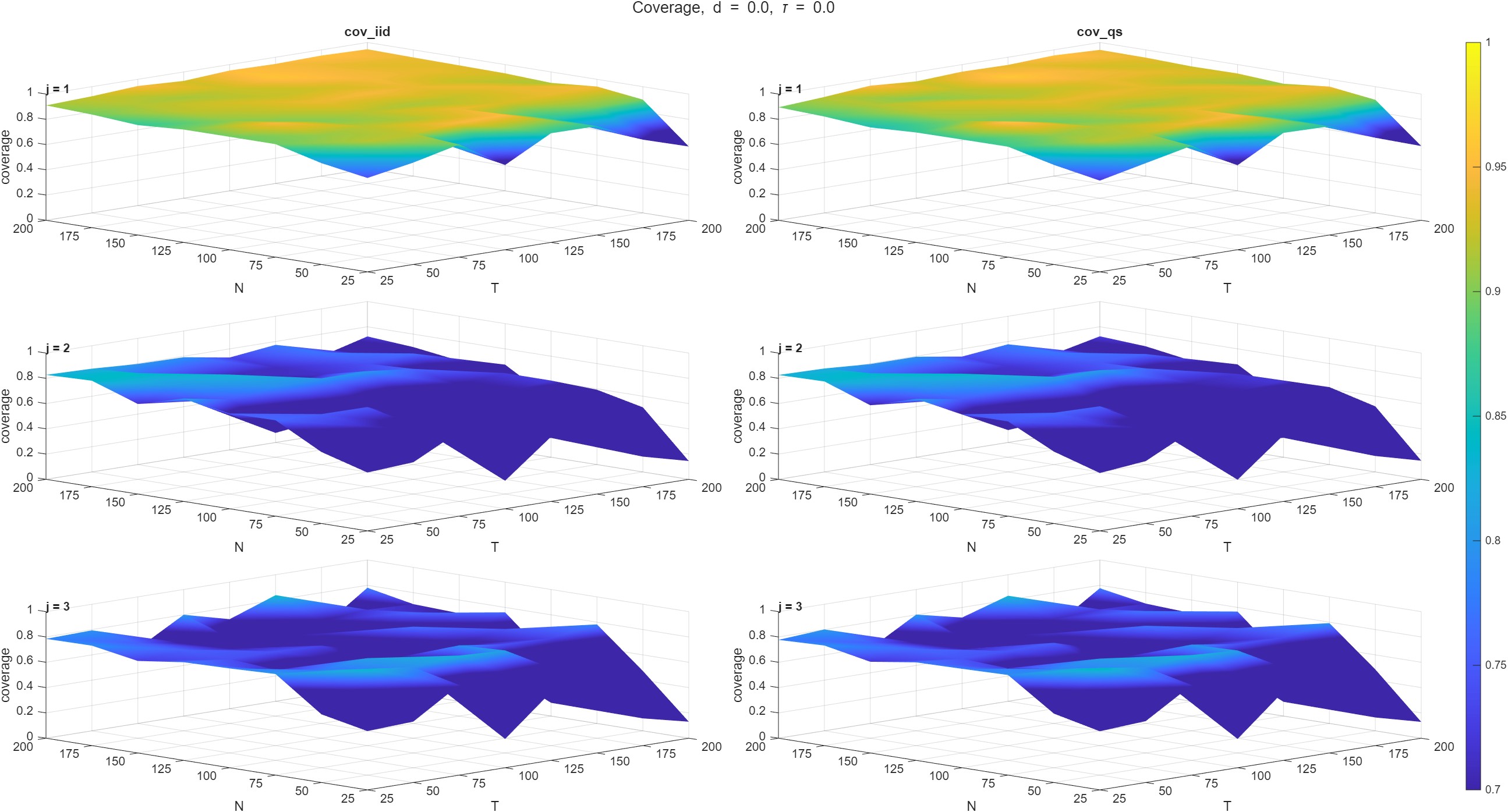}
\includegraphics[trim=0 0 100 80, clip, width=0.55\textwidth]{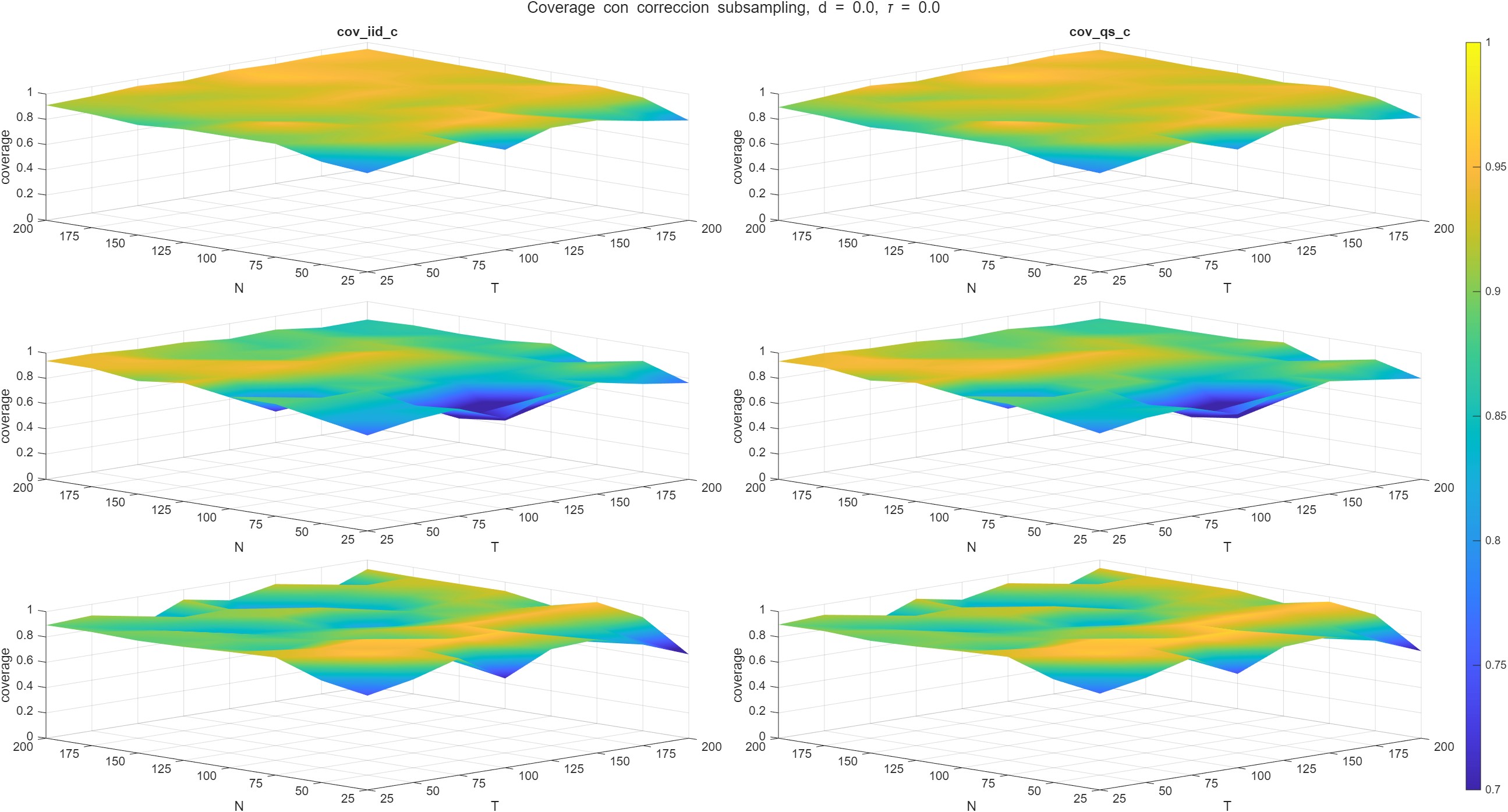}
\caption{Empirical coverages of confidence intervals for the first loading of each of the factors for the DFM with $r=3$ and $\phi=0.7$ with cross-sectionally uncorrelated white noise idiosyncratic components with the signal-to-noise ratio is 1. The loading of the first factor in first row, of the second factor in second row, and of the third factor in third row. The MSE of the loadings are obtained without (left panel) and with subsampling correction (right panel). Within each panel, the MSE are obtained assuming serially uncorrelated idiosyncratic components (first column) using the HAC estimator with HAR inference (second column).}
\label{fig:Coverage_R3}
\end{figure}

Finally, to analyse the effect on inference on PC loadings of estimating more than one factor, we consider the DFM with $r=3$ described above. Figure \ref{fig:Coverage_R3} plots the empirical coverages of the 95\% confidence interval for the estimated loadings of each of the three factors on the first variable in $Y_t$, both in the case of idiosyncratic serial and cross-sectional uncorrelation, and for serially correlated idiosyncratic components with $\delta=0.5$. The confidence intervals are constructed assuming normality with the covariance matrix of the loadings estimated by $\hat{\Omega}^{(0)}$ and using the HAR critical values with the covariance matrix estimated by $\hat{\Omega}^{(QS)}$. Once more, we can observe a clear deterioration of the coverages which can be under 0.7 when $N$ is not large enough.

\begin{figure}[h!]
\includegraphics[trim=0 0 0 80, clip, width=1.0\textwidth]{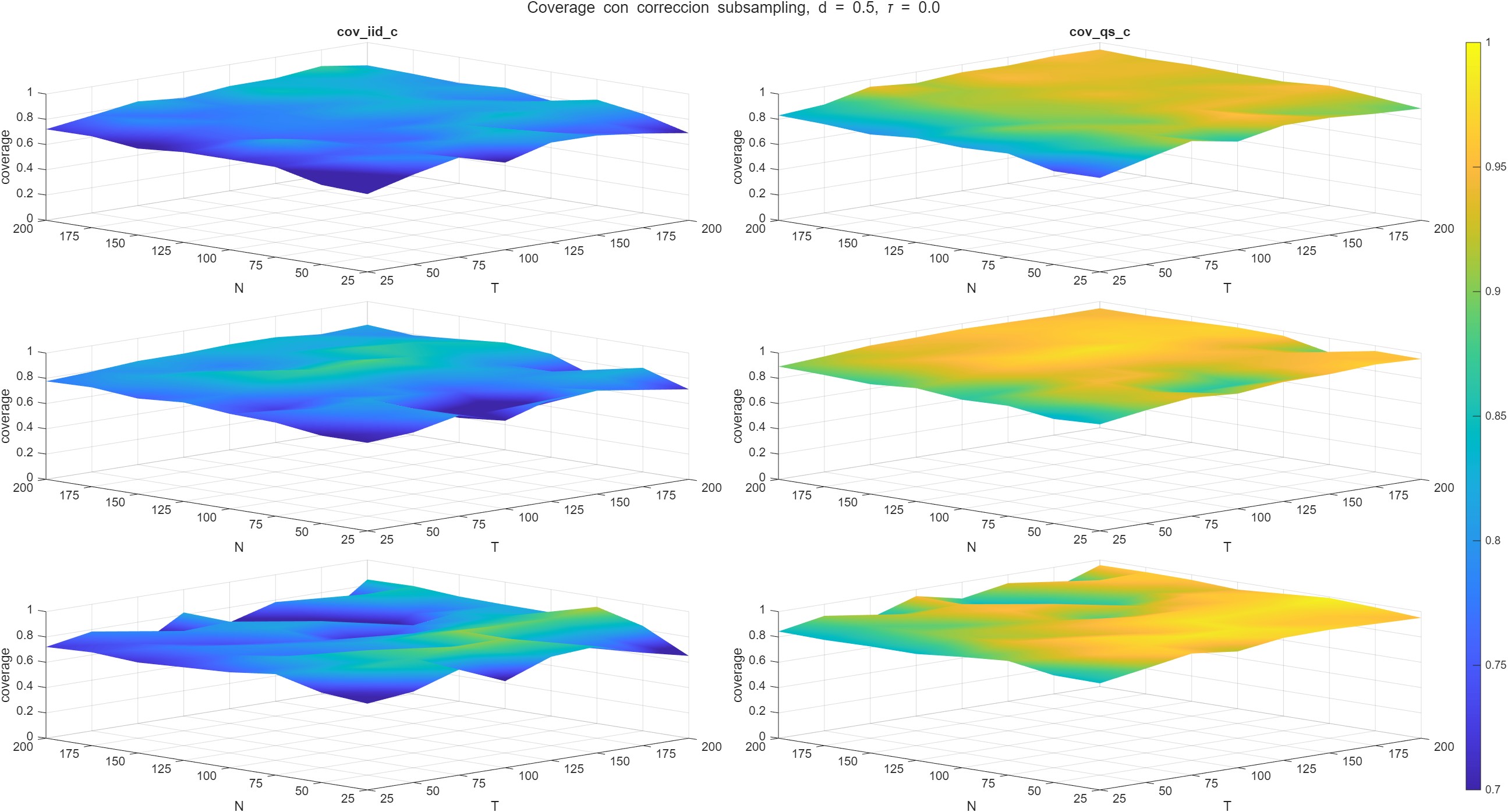}
\caption{Empirical coverages of confidence intervals for the first loading of each of the factors for the DFM with $r=3$ and $\phi=0.7$ with cross-sectional uncorrelated AR(1) idiosyncratic components with $\delta=0.5$ and with unity signal-to-noise ratio. Coverages of loadings of the first factor in first row, of the second factor in second row, and of the third factor in third row. The MSE of the loadings are obtained with subsampling correction, assuming serially uncorrelated idiosyncratic components (first column) and using the HAC estimator with HAR inference (second column).}
\label{fig:Coverage_R3_2}
\end{figure}

 \subsection{Properties of significance tests}

In this subsection, we carry out Monte Carlo experiments to study the performance of test the individual significance of the loading corresponding to the first variable in the system, $H_0:\lambda_1=0$ in the DFM described above with $r=1$. Table \ref{tab:Power_1} reports the size of the significance test when the idiosyncratic components are generated by AR(1) processes with $\delta=0$ and 0.5 and inference is based on the asymptotic normal distribution with the covariance matrix of the loadings estimated assuming that the idiosyncratic components are uncorrelated. Table \ref{tab:Power_2} also reports sizes when the covariance matrix is computed as proposed in this paper using $\Omega^{(QS)}$ corrected by subsampling and inference is carried out with the HAR critical values. 

\begin{table}[h!]
%{\scriptsize
\centering
\begin{spacing}{0.75}
\begin{tabular}{c|ccc|ccc|ccc}
\hline 
\textbf{\scriptsize{}$N$} &  & \textbf{\scriptsize{}50} &  &  & \textbf{\scriptsize{}100} &  &  & \textbf{\scriptsize{}200} & \tabularnewline
\hline 
\textbf{\scriptsize{}$T$} & \textbf{\scriptsize{}50} & \textbf{\scriptsize{}100} & \textbf{\scriptsize{}200} & \textbf{\scriptsize{}50} & \textbf{\scriptsize{}100} & \textbf{\scriptsize{}200} & \textbf{\scriptsize{}50} & \textbf{\scriptsize{}100} & \textbf{\scriptsize{}200}\tabularnewline
\hline
\multicolumn{10}{c}{\textbf{\scriptsize{}Size $\lambda_1=0$}}\tabularnewline
\hline
\textbf{\scriptsize{}$\delta=0$} & {\scriptsize{}0.10} & {\scriptsize{}0.11} & {\scriptsize{}0.08} & {\scriptsize{}0.10} & {\scriptsize{}0.08} & {\scriptsize{}0.08} & {\scriptsize{}0.07} & {\scriptsize{}0.06} & {\scriptsize{}0.06}\tabularnewline
\textbf{\scriptsize{}$\delta=0.5$} & {\scriptsize{}0.24} & {\scriptsize{}0.28} & {\scriptsize{}0.22} & {\scriptsize{}0.22} & {\scriptsize{}0.23} & {\scriptsize{}0.24} & {\scriptsize{}0.18} & {\scriptsize{}0.22} & {\scriptsize{}0.22}\tabularnewline
\hline
\multicolumn{10}{c}{\textbf{\scriptsize{}Power}}\tabularnewline
\hline
\textbf{\scriptsize{}$\lambda_1=0.5$} &  &  &  &  &  &  &  &  & \tabularnewline
\hline
\textbf{\scriptsize{}$\delta=0$} & {\scriptsize{}0.41} & {\scriptsize{}0.47} & {\scriptsize{}0.92} & {\scriptsize{}0.30} & {\scriptsize{}0.51} & {\scriptsize{}0.67} & {\scriptsize{}0.62} & {\scriptsize{}1.00} & {\scriptsize{}0.84} \tabularnewline
\textbf{\scriptsize{}$\delta=0.5$} & {\scriptsize{}0.46} & {\scriptsize{}0.50} & {\scriptsize{}0.82} & {\scriptsize{}0.37} & {\scriptsize{}0.52} & {\scriptsize{}0.62} & {\scriptsize{}0.59} & {\scriptsize{}0.99} & {\scriptsize{}0.75} \tabularnewline 
\hline 
\textbf{\scriptsize{}$\lambda_1=1$} &  &  &  &  &  &  &  &  & \tabularnewline
\hline
\textbf{\scriptsize{}$\delta=0$} & {\scriptsize{}0.90} & {\scriptsize{}0.93} & {\scriptsize{}1.00} & {\scriptsize{}0.72} & {\scriptsize{}0.97} & {\scriptsize{}1.00} & {\scriptsize{}0.99} & {\scriptsize{}1.00} & {\scriptsize{}1.00} \tabularnewline
\textbf{\scriptsize{}$\delta=0.5$} & {\scriptsize{}0.81} & {\scriptsize{}0.85} & {\scriptsize{}1.00} & {\scriptsize{}0.66} & {\scriptsize{}0.90} & {\scriptsize{}0.97} & {\scriptsize{}0.97} & {\scriptsize{}1.00} & {\scriptsize{}1.00} \tabularnewline 
\hline 
\textbf{\scriptsize{}$\lambda_1=2$} &  &  &  &  &  &  &  &  & \tabularnewline
\hline
\textbf{\scriptsize{}$\delta=0$} & {\scriptsize{}1.00} & {\scriptsize{}1.00} & {\scriptsize{}1.00} & {\scriptsize{}1.00} & {\scriptsize{}1.00} & {\scriptsize{}1.00} & {\scriptsize{}1.00} & {\scriptsize{}1.00} & {\scriptsize{}1.00} \tabularnewline
\textbf{\scriptsize{}$\delta=0.5$} & {\scriptsize{}1.00} & {\scriptsize{}1.00} & {\scriptsize{}1.00} & {\scriptsize{}0.98} & {\scriptsize{}1.00} & {\scriptsize{}1.00} & {\scriptsize{}1.00} & {\scriptsize{}1.00} & {\scriptsize{}1.00} \tabularnewline 
\hline
\end{tabular}
\end{spacing}
%}
\caption{Monte Carlo size and power of significance test $H_0: \lambda_1=0$ with empirical size 5\% using the asymptotic (normal) distribution with the covariance matrix estimated using $\widehat{\Omega}^{(0)}$, when the idiosyncratic noises are AR(1) processes with parameter $\delta$.}
\label{tab:Power_1}
\end{table}

\begin{table}[h!]
%{\scriptsize
\centering
\begin{spacing}{0.75}
\begin{tabular}{c|ccc|ccc|ccc}
\hline 
\textbf{\scriptsize{}$N$} &  & \textbf{\scriptsize{}50} &  &  & \textbf{\scriptsize{}100} &  &  & \textbf{\scriptsize{}200} & \tabularnewline
\hline 
\textbf{\scriptsize{}$T$} & \textbf{\scriptsize{}50} & \textbf{\scriptsize{}100} & \textbf{\scriptsize{}200} & \textbf{\scriptsize{}50} & \textbf{\scriptsize{}100} & \textbf{\scriptsize{}200} & \textbf{\scriptsize{}50} & \textbf{\scriptsize{}100} & \textbf{\scriptsize{}200}\tabularnewline
\hline
\multicolumn{10}{c}{\textbf{\scriptsize{}Size $\lambda_1=0$}}\tabularnewline
\hline
\textbf{\scriptsize{}$\delta=0$} & {\scriptsize{}0.07} & {\scriptsize{}0.08} & {\scriptsize{}0.05} & {\scriptsize{}0.07} & {\scriptsize{}0.06} & {\scriptsize{}0.06} & {\scriptsize{}0.05} & {\scriptsize{}0.05} & {\scriptsize{}0.05}\tabularnewline
\textbf{\scriptsize{}$\delta=0.5$} & {\scriptsize{}0.12} & {\scriptsize{}0.12} & {\scriptsize{}0.08} & {\scriptsize{}0.12} & {\scriptsize{}0.10} & {\scriptsize{}0.08} & {\scriptsize{}0.10} & {\scriptsize{}0.10} & {\scriptsize{}0.07}\tabularnewline
\hline
\multicolumn{10}{c}{\textbf{\scriptsize{}Power}}\tabularnewline
\hline
\textbf{\scriptsize{}$\lambda_1=0.5$} &  &  &  &  &  &  &  &  & \tabularnewline
\hline
\textbf{\scriptsize{}$\delta=0$} & {\scriptsize{}0.34} & {\scriptsize{}0.39} & {\scriptsize{}0.88} & {\scriptsize{}0.24} & {\scriptsize{}0.45} & {\scriptsize{}0.62} & {\scriptsize{}0.55} & {\scriptsize{}1.00} & {\scriptsize{}0.80} \tabularnewline
\textbf{\scriptsize{}$\delta=0.5$} & {\scriptsize{}0.30} & {\scriptsize{}0.31} & {\scriptsize{}0.59} & {\scriptsize{}0.23} & {\scriptsize{}0.33} & {\scriptsize{}0.37} & {\scriptsize{}0.44} & {\scriptsize{}0.93} & {\scriptsize{}0.48} \tabularnewline 
\hline 
\textbf{\scriptsize{}$\lambda_1=1$} &  &  &  &  &  &  &  &  & \tabularnewline
\hline
\textbf{\scriptsize{}$\delta=0$} & {\scriptsize{}0.85} & {\scriptsize{}0.89} & {\scriptsize{}1.00} & {\scriptsize{}0.66} & {\scriptsize{}0.95} & {\scriptsize{}1.00} & {\scriptsize{}0.99} & {\scriptsize{}1.00} & {\scriptsize{}1.00} \tabularnewline
\textbf{\scriptsize{}$\delta=0.5$} & {\scriptsize{}0.69} & {\scriptsize{}0.68} & {\scriptsize{}0.99} & {\scriptsize{}0.53} & {\scriptsize{}0.76} & {\scriptsize{}0.87} & {\scriptsize{}0.91} & {\scriptsize{}1.00} & {\scriptsize{}0.97} \tabularnewline 
\hline 
\textbf{\scriptsize{}$\lambda_1=2$} &  &  &  &  &  &  &  &  & \tabularnewline
\hline
\textbf{\scriptsize{}$\delta=0$} & {\scriptsize{}1.00} & {\scriptsize{}1.00} & {\scriptsize{}1.00} & {\scriptsize{}0.99} & {\scriptsize{}1.00} & {\scriptsize{}1.00} & {\scriptsize{}1.00} & {\scriptsize{}1.00} & {\scriptsize{}1.00} \tabularnewline
\textbf{\scriptsize{}$\delta=0.5$} & {\scriptsize{}1.00} & {\scriptsize{}0.99} & {\scriptsize{}1.00} & {\scriptsize{}0.95} & {\scriptsize{}1.00} & {\scriptsize{}1.00} & {\scriptsize{}1.00} & {\scriptsize{}1.00} & {\scriptsize{}1.00} \tabularnewline 
\hline
\end{tabular}
\end{spacing}
%}
\caption{Monte Carlo size and power of significance test $H_0: \lambda_1=0$ with empirical size 5\% using HAR inference with the covariance matrix estimated using $\widehat{\Omega}^{(QS)}$, when the idiosyncratic noises are AR(1) processes with parameter $\delta$.}
\label{tab:Power_2}
\end{table}

The conclusions about size are obviously in concordance with those obtained above when analysing the empirical densities and the coverages of the confidence intervals. If the idiosyncratic components are white noise, the sizes are similar regardless of whether $\widehat{\Omega}^{(0)}$ or $\widehat{\Omega}^{(QS)}$ are used to estimate the covariance of the loadings and of whether normal or HAR critical values are used when testing. If $N$ is not very large, the significance tests are slightly oversized for small $T$. However, if $T$ is large enough, the size is close to the nominal 5\%. If the idiosyncratic components are serially correlated, using $\widehat{\Omega}^{(0)}$ with normal critical values always leads to severe oversized significance tests. This distortion is clearly corrected by using the proposed $\widehat{\Omega}^{(QS)}$ estimator of the covariance of the loadings corrected by subsampling and with HAR critical values.

Tables \ref{tab:Power_1} and \ref{tab:Power_2} also report the power of the significance tests when $\lambda_1=0.5$, 1 and 2. Note that in the first case, the signal to noise is very small with the variability of the idiosyncratic component being four times that of the common components. The signal to noise ratio is 1 when $\lambda_1=1$, and 4 when $\lambda_1=2$. When looking at power, we can observe that the inference for the loadings proposed in this paper leads to significance tests with good power if the signal of the common component is strong enough. Even if the signal is very weak, i.e. $\lambda_1=0.5$, there is some non-negligible power. Finally, note that when the idiosyncratic components are truly uncorrelated, there is not lost of power when using the proposed inference instead of the traditional one based on $\widehat{\Omega}^{(0)}$ and normal critical values. 

%\subsection{Joint tests on loadings across several units}

\section{Empirical application}
\label{sec:empirics}

Analysing economic convergence in the US has been the subject of research in a large number of works; see, for example, Carlino and Mills (1996), Tomljanovich and Vogelsang (2001), Carvalho and Harvey (2005)%\footnote{Fit multivariate unobserved component time series models to annual data on real income per capita observed from 1950 to 1999, to analyse stylised facts about cycles and convergence and conclude that there appear to be convergence of all regions apart from the two richest which are diverging from the other six regions as well as from each other.}
, Choi and Wang (2015) and Miles (2020). In this section, we illustrate the relevance of using the estimator of the PC loadings proposed in this paper when looking for underlying factors that can be common to particular groups of states. 

Quarterly data on real \textit{per capita} GDP for each of the 50 states plus Columbia district is observed quarterly from 2005Q1 to 2025Q4. The data were obtained from the US Bureau of Economic Analysis Regional Data and deflated by the US implicit price deflator (2017=100). Denote by $y_{it}=\log \left( \frac{GDP_{it}}{GDP_{US,t} }\right)$, where $GDP_{it}$ is the \textit{per capita} GDP of state $i=1,...,51$, at time $t=1,...,84$, and $GDP_{US,t}$ is the \textit{per capita} GDP of the US at time $t$. This transformation is often used in the empirical analysis of regional convergence; see, for example, Barro and Sala-i-Martin (1991, 1992), who analyse whether regional economies tend to be closer in terms of \textit{per capita} product or income. Defining the \textit{per capita} GDP of each state with respect to the aggregated \textit{per capita} GDP in the US, the common component associated with US growth is eliminated and the analysis is focused on the relative position of each state with respect to the national benchmark; see Carvalho and Harvey (2005), who show the relevance of accounting for trends and cycles when dealing with convergence.  Furthermore, $y_{it}$ can be interpreted as a percentage relative deviation, which is particularly appropriate when the goal is not just the analysis of convergence but also to identify divergence and the presence of divergence clubs; see, for example, Quah (1996, 1997) and Phillips and Sul (2007). 

\begin{figure}[h!]
\includegraphics[width=1.0\textwidth]{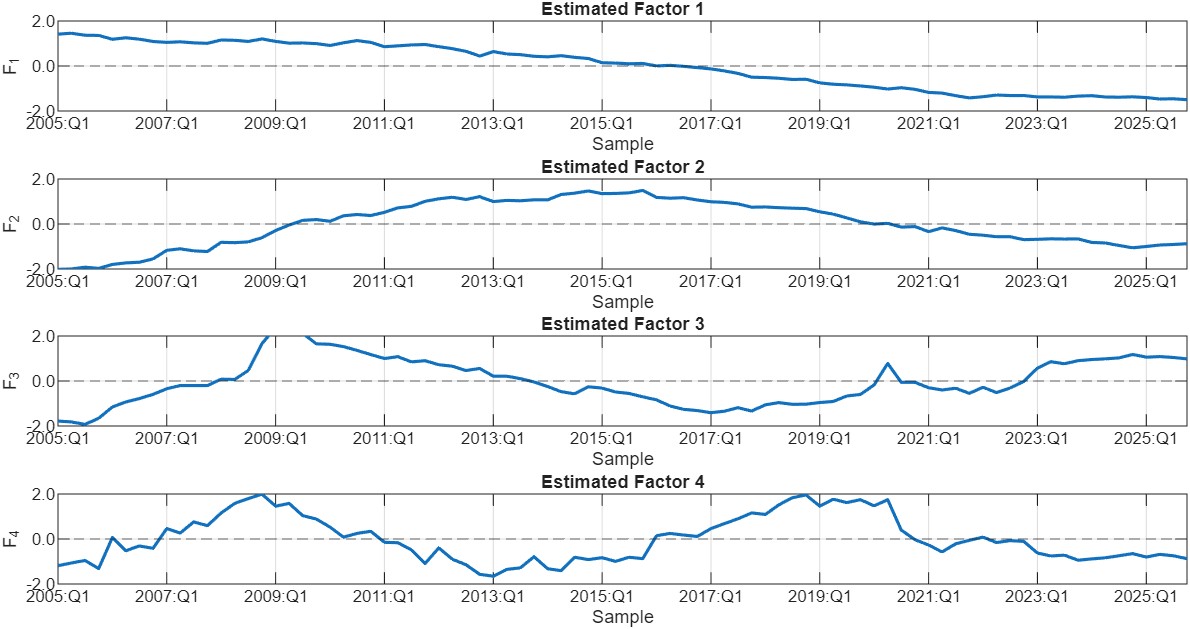}
\caption{Estimated common factors of percentage relative \textit{per capita} GDP.}
\label{fig:Factors_empiric}
\end{figure}

After standardizing the data, the number of factors is determined using the criteria by Alessi, Barigozzi and Capasso (2010), which determine either $r=3$ or $r=5$. Consequently, and using also the information in the corresponding screeplot, we decide to extract $r=4$ factors; see Appendix B in which we show that $r=4$ factors explain 90\% of the variability in the data set.  The factors, which should be interpreted as common movements in each state with respect to its relative position with respect the US, are extracted by PC and plotted in Figure \ref{fig:Factors_empiric}. The first factor shows a clear negative trend with an evolving slope, which is smooth until the first global financial crisis in 2010, increases after, and becomes flat after the COVID19 pandemic in 2022. The profile of the second factor is very  different, showing a positive slope until 2012, which becomes negative after 2016. As in the first factor, the slope seems to become flat after 2023. The third and fourth factors are more erratic representing mid-term cycles. It is remarkable that these factors are also approximately flat since 2023. Therefore, the evolution of all four factors is approximately flat since 2023, implying that, if the idiosyncratic components are stationary, whatever the relative position of each state with respect to the aggregate US \textit{per capita} GDP, this position has not changed during the last three years of the sample period, from 2023 to 2025. Finally, note that the idiosyncratic components may be assumed to be weakly stationary with temporal dependence; see Appendix B, which reports $p$-values of the Dickey-Fuller test of non-stationarity together with heat-maps of the autocorrelations of the idiosyncratic components.

\begin{figure}[h!]
\includegraphics[trim=0 0 0 0, clip, width=1.0\textwidth]{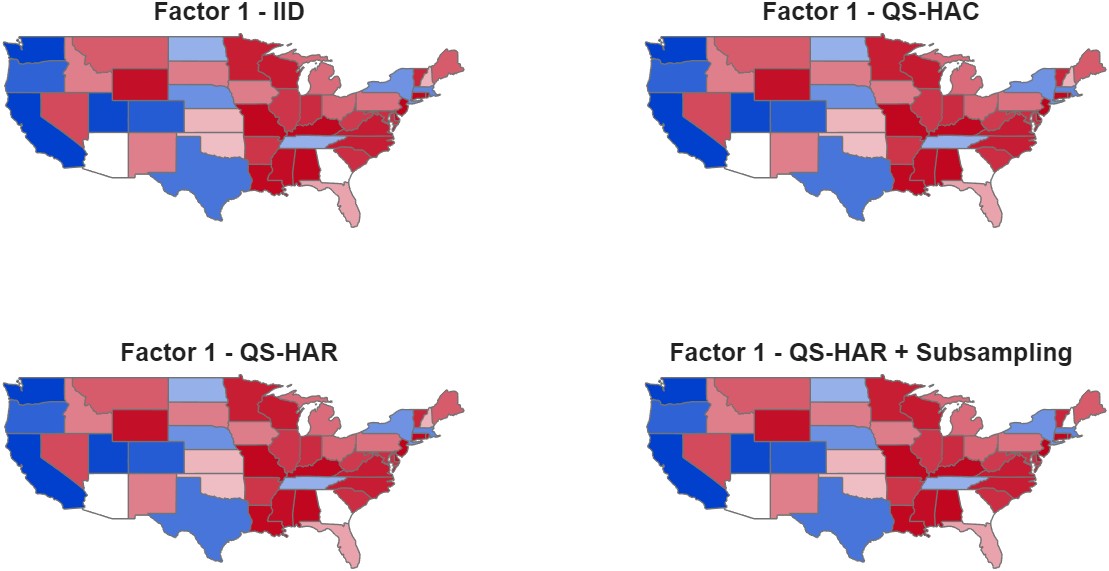}
\includegraphics[trim=0 0 0 28, clip, width=1.0\textwidth]{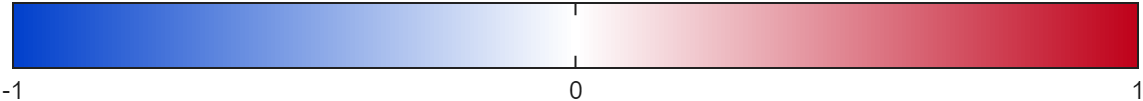}
\caption{Map of US states with significant loadings of first PC factor when the test is based on normal critical values and the covariance matrix $\Omega$ estimated using: i) $\hat{\Omega}^{(0)}$ (top left panel); ii) $\hat{\Omega}^{(QS)}$ (top right panel), and with HAR critical values and the covariance matrix estimated with $\hat{\Omega}^{(QS)}$ without (bottom left panel) and with (bottom right panel) subsampling correction.}
\label{fig:map1_empiric}
\end{figure}

\begin{figure}[h!]
\includegraphics[trim=0 0 0 0, clip, width=1.0\textwidth]{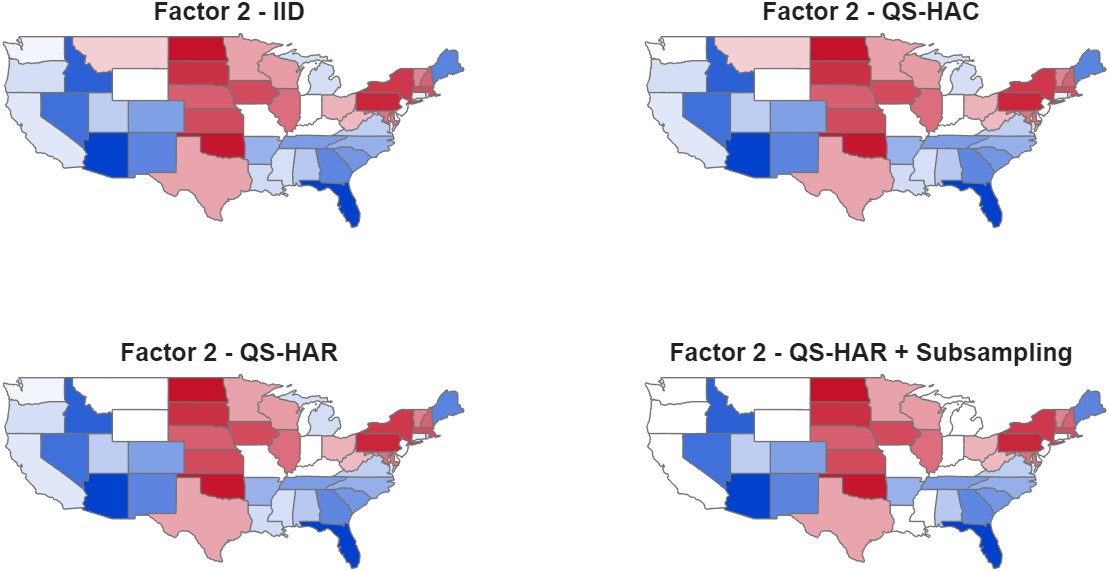}
\includegraphics[trim=0 0 0 28, clip, width=1.0\textwidth]{paleta_colores.png}
\caption{Map of US states with significant loadings of second factor when the test is based on normal critical values and the covariance matrix $\Omega$ estimated using: i) $\hat{\Omega}^{(0)}$ (top left panel); ii) $\hat{\Omega}^{(QS)}$ (top right panel), and with HAR critical values and the covariance matrix estimated with $\hat{\Omega}^{(QS)}$ without (bottom left panel) and with (bottom right panel) subsampling correction.}
\label{fig:map2_empiric}
\end{figure}

\begin{figure}[h!]
\includegraphics[trim=0 0 0 0, clip, width=1.0\textwidth]{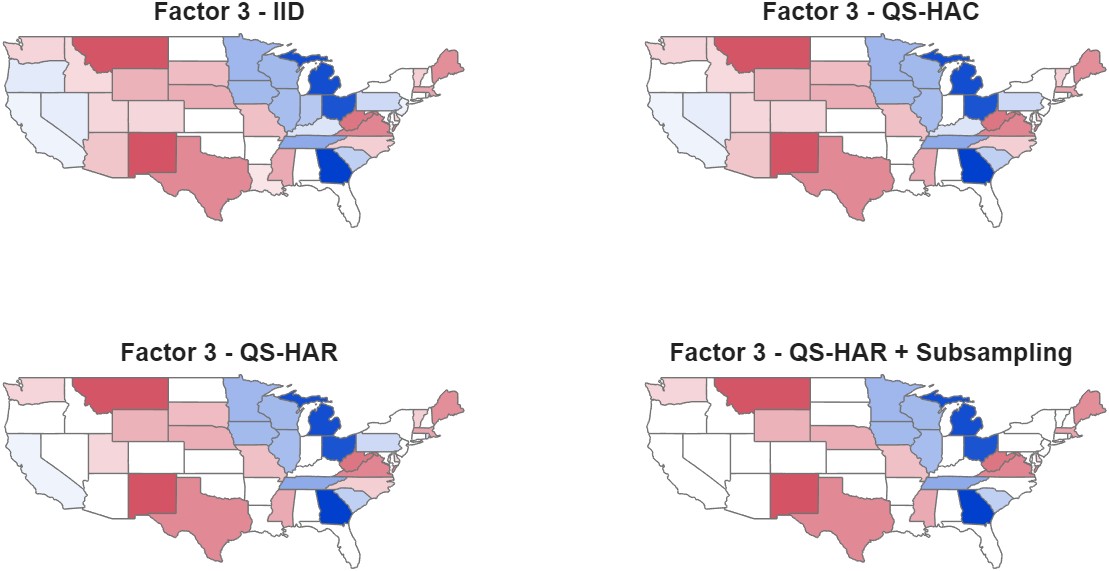}
\includegraphics[trim=0 0 0 28, clip, width=1.0\textwidth]{paleta_colores.png}
\caption{Map of US states with significant loadings of third factor when the test is based on normal critical values and the covariance matrix $\Omega$ estimated using: i) $\hat{\Omega}^{(0)}$ (top left panel); ii) $\hat{\Omega}^{(QS)}$ (top right panel), and with HAR critical values and the covariance matrix estimated with $\hat{\Omega}^{(QS)}$ without (bottom left panel) and with (bottom right panel) subsampling correction.}
\label{fig:map3_empiric}
\end{figure}

\begin{figure}[h!]
\includegraphics[trim=0 0 0 0, clip, width=1.0\textwidth]{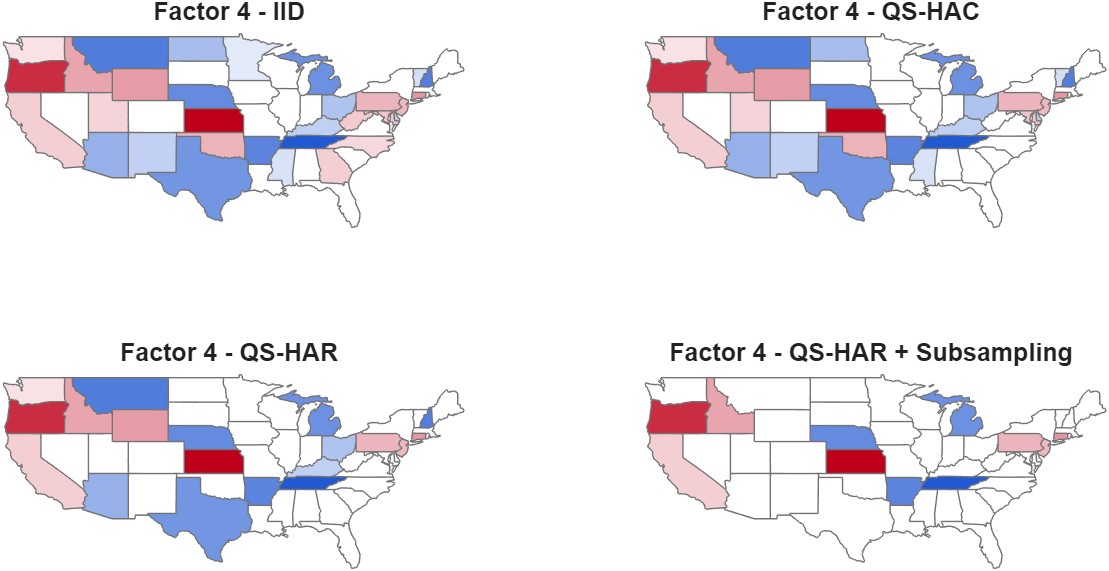}
\includegraphics[trim=0 0 0 28, clip, width=1.0\textwidth]{paleta_colores.png}
\caption{Map of US states with significant loadings of fourth factor when the test is based on normal critical values and the covariance matrix $\Omega$ estimated using: i) $\hat{\Omega}^{(0)}$ (top left panel); ii) $\hat{\Omega}^{(QS)}$ (top right panel), and with HAR critical values and the covariance matrix estimated with $\hat{\Omega}^{(QS)}$ without (bottom left panel) and with (bottom right panel) subsampling correction.}
\label{fig:map4_empiric}
\end{figure}

Recall that the estimated factors plotted in Figure \ref{fig:Factors_empiric} represent the common movements of the states relative to the national GDP. To properly interpret these common movements, we need to analyse the loadings, which represent the exposure of each state to these common movements. To help with interpretation of the factors, Figures \ref{fig:map1_empiric} to \ref{fig:map4_empiric} plot colour-maps of the US states representing the loadings estimated for each of the four PC factors, respectively. The colours in these figures represent whether the estimated loadings are significant or not and, in the former case, the sign and magnitude of the estimated loadings. The significance of the estimated loadings is tested estimating their covariance matrix using $\hat{\Omega}^{(0)}_i$ and $\hat{\Omega}^{(QS)}_i$, for $i=1,...,51$, with the critical values obtained from the normal distribution. For the latter estimator of the covariance matrix, we chose $q=\lfloor 0.17 Y \rfloor$ and also consider HAR critical values without and with the subsampling correction. The estimated PC loadings together with 95\% confidence intervals obtained  are reported in Appendix B. The estimated loadings of the first factor, represented in Figure \ref{fig:map1_empiric}, separate the states into two groups with opposite dynamics. Most loadings are significant, with the states with positive loadings (Alabama, Arkansas, Connecticut, Kentucky, Louisiana, Mississippi, Missouri, New Jersey, Rhode Island, Virginia, Wisconsin and Wyoming) worsening their relative position with respect to the national \textit{per capita} GDP. On the contrary, the states with negative loadings (California, Colorado, Massachusetts, New York, Oregon, Texas, Utah and Washington) have a relatively better position with respect to the national \textit{per capita} GDP. The first factor divides the states into a group of occidental and metropolitan states, including the West Coast, part of Mountain West, Texas and advanced states of the North-east, and a larger group of states in the South, Midwest and North-Easter, characterized by more traditional production structures; see Giannone (2022), who argues that skill-biased technical change weakened regional convergence by disproportionately benefiting locations with higher concentrations of skilled labour. In our setting, the first factor appears to capture a division between technology- and advanced-service-intensive states and states with more traditional production structures. Figure \ref{fig:map1_empiric} also shows that this patter is relatively robust to the particular estimator of the covariance matrix of the loadings used to test for significance.

Figure \ref{fig:map2_empiric} represents the colour-map of the states with significant loadings with respect to the second factor. The states with positive loadings are Illinois, Iowa, Kansas, Maryland, Massachusetts, Nebraska, New York, North Dakota, Oklahoma, Pennsylvania, South Dakota, Texas, Vermont and Wisconsin, while Arizona, Florida, Georgia, Idaho, Nevada, North Carolina, South Carolina, Tennessee (Sun Belt/Southeast/West interior) and other states in Plains-Midwest-North/Easter have negative loadings. As with the first factor, this division is nearly the same regardless of the particular estimator of the covariance matrix of the loadings used to test their significance. The second factor is better interpreted through the lens of structural transformation and geographically uneven exposure to sectoral shocks. Consistent with Kim and Lee (2024), structural change has been central to US regional convergence, although its traditional farm-to-nonfarm channel has become much less relevant after 1990. In our setting, the factor seems to capture more recent forms of structural differentiation across states. In particular, it distinguishes states with older manufacturing, energy, aggro-industrial or productive structures, such as Ohio, Michigan, Wisconsin, Iowa, Illinois, Pennsylvania, Kansas, Oklahoma, North Dakota and Texas, from faster-growing Sun Belt economies more exposed to construction, real estate, logistics and population-driven local services, such as Florida, Arizona, Nevada, Georgia, North Carolina, South Carolina and Tennessee. This interpretation is also consistent with Autor, Dorn and Hanson (2013, 2015), who show that local exposure to trade and technological shocks depends strongly on initial industrial and occupational structure, and with Moretti (2012), who emphasizes the increasing spatial differentiation of local economies and employment opportunities.

Figure \ref{fig:map3_empiric} represents the colour-map of the states with significant  loadings for the third factor. In this case, there is a different pattern of significant loadings depending on the particular estimator of the covariance matrix of the loadings used to test their significance. When the covariance matrix is estimated using $\widehat{\Omega}^{(0)}_i$, i.e. assuming that the idiosyncratic components are serially uncorrelated, a large number of states have significant loadings, without a clear pattern of significant/nonsignificant states. However, if the covariance matrix is estimated using $\widehat{\Omega}^{(QS)}_i$ corrected with subsampling and with inference carried out using the HAR critical values, there is a clearer pattern of significant states, illustrating the difference in interpretation of the factors depending on the particular estimator/inference used to estimate the covariance matrix of the loadings. In particular, when using HAR inference with the covariance matrix estimated using $\widehat{\Omega}^{(QS)}_i$ corrected with subsampling, the significant positive loadings correspond to Alaska, Montana, New Mexico, Texas, West Virginia and Wyoming, while negative loadings are associated to Georgia, Illinois, Iowa, Michigan, Minnesota, Ohio, South Carolina, Tennessee and Wisconsin. The production of the former states is associated to natural resources and energy, while the latter are located in the Midwest/Great Lakes area with strong industrial and manufacturing sectors.

Differences between significant loadings depending on the particular estimator/inference used to test for significance are even more evident in Figure \ref{fig:map4_empiric}, which represents the significant loadings associated with the fourth factor. The number of significant loadings is relatively small when using HAR inference and estimating the covariance matrix  using $\hat{\Omega}^{(QS)}_i$ corrected by subsampling. The states with positive loadings are California, Connecticut, Idaho, Kansas, New Jersey, Oregon and Pennsylvania, while the loadings are negative in Arkansas, Michigan, Nebraska and Tennessee. The states with positive loadings are among those with largest contribution to national DGP. In particular, the largest contributor is California, followed by New York (note that New Jersey and Connecticut are part of the metropolitan area of New York). Furthermore, the technological industry of Oregon is strongly connected to California, acting as a natural location for data centres and technological companies that want to move out of California. Finally, note that the four states with negative loadings are linked through major supply chains, inland waterway systems, and trade corridors. Michigan is the historic centre of the US auto industry and relies heavily on automotive and parts suppliers located across Tennessee and Arkansas, Furthermore, the Interstate 69 Corridor physically links Michigan down through Tennessee and Arkansas. Finally, the Mississippi River acts as a primary artery for agricultural exports shipped up from Arkansas to Midwestern states, including Nebraska's farming.

The analysis above shows that if the significance analysis is carried out using test computed with the covariance matrix of the loadings estimated using $\widehat{\Omega}^{(0)}_i$, the number of significant loadings is larger and the interpretation of the factors is more diffuse, mixing factors that affect all states with those that load only on particular groups of states. However, computing the covariance matrix using $\widehat{\Omega}^{(QS)}_i$ corrected with subsampling and inference being based on HAR critical values, allows to separate more neatly the factors loading on all states from those loadings on groups of states. From the point of view of convergence, the estimated ML-DFM shows that during 2005 to 2025, the deviations of each state with respect to its relative position with respect to the national \textit{per capita} GDP have common persistent components and heterogeneous regional cycles. The loadings allow the identification of this heterogeneity with some coherent economic regions: i) an occidental/metropolitan region as opposite to region with more traditional production; ii) the Sun Belt/Southeast region as opposite to the Plains-Midwest-Northeast region; iii) the manufacturing Midwest region as opposite to states with natural resources and energy; and iv) regions associate with trading channels.

\section{Conclusions}
\label{sec:conclusions}

In this paper, we analyse the performance of inference on PC estimates of the loadings in approximate DFMs when it is based on the asymptotic distribution. Extensive simulations show that traditional inference based on estimating the asymptotic covariance of the loadings assuming that the idiosyncratic components are serially uncorrelated and using the critical values of the normal distribution leads to overrejection of the null in significance tests or to confidence intervals with coverages well under the nominal. Using the Bartlett kernel to estimate the covariance matrix of the loadings taking into account serially idiosyncratic dependence, helps but inference still has poor properties and the interpretation of the factors can be seriously  blurred. We propose estimating the covariance matrix of PC loadings using the more efficient QS kernel and correcting it with subsampling to take into account the uncertainty associated to the estimation of the factors, which can be relevant when the cross-sectional dimension is not very large. Furthermore, as usual in the related literature for regression models with autocorrelated errors, we propose using fixed-b critical values (HAR inference) instead of normal critical values for inference.  We show that the size of significance tests and the coverages of the confidence intervals for PC loadings are, in this case, closer to nominal values.

In an empirical illustration to analyse convergence patterns in the US states, we show that using HAR inference with the covariance matrix estimated using a QS kernel corrected with subsampling, allows for a more precise measurement of the uncertainty associated to the estimation of PC loadings and, in doing so, for a more neat identification of the economic regions that share common movements.

\theendnotes

\newpage

\section*{Appendix A: Further Monte Carlo results}

\setcounter{figure}{0}
\renewcommand{\thefigure}{A.\arabic{figure}}

\setcounter{secnumdepth}{0}

In this Appendix, we report the results of further Monte Carlo experiments. In particular, we report results for idiosyncratic components with cross-sectional correlation, conditional heteroscedasticity, and Student-7 distribution. 

\subsection{Cross-correlated idiosyncratic components}

\begin{figure}[h!]
\includegraphics[trim=0 0 0 62, clip, width=1.0\textwidth]{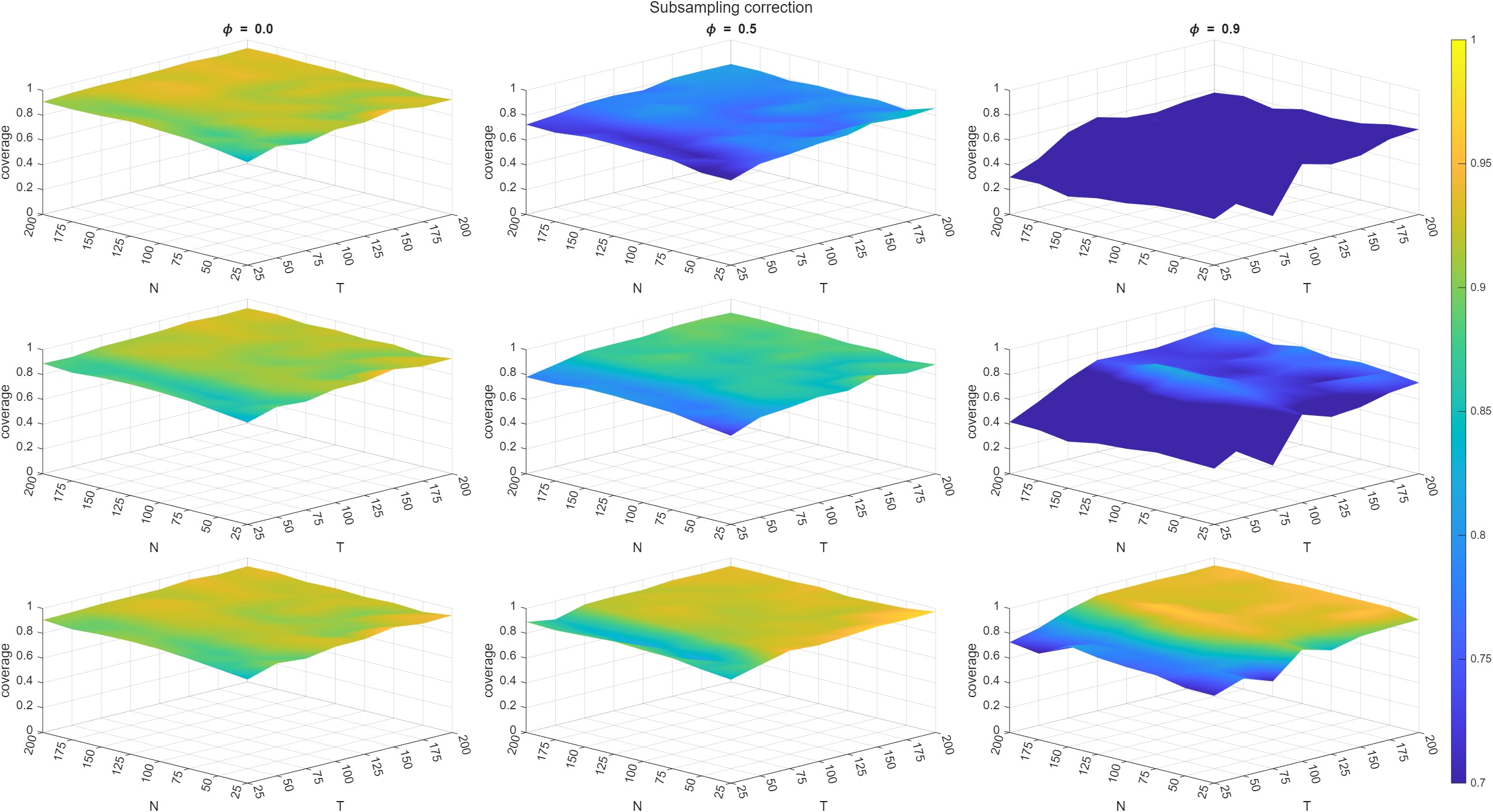}
\vspace{-2mm}
\caption{Empirical coverages of point-wise confidence intervals for $\lambda_1$ for the DFM with $r=1$, $\phi=0.7$ and $c=1$, with idiosyncratic components being cross-sectionally dependent with $\tau=-0.5$: serially uncorrelated (first column); serially dependent with $\delta=0.5$ (second column); and serially dependent with $\delta=0.9$ (third column). The asymptotic covariance matrix is computed with subsampling correction without (first row) and with the HAC correction, and with inference based on normality (second row) and on HAR (third row).}
\label{fig:Coverage_2}
\end{figure}

Figure \ref{fig:Coverage_2} plots the empirical coverages when the idiosyncratic components are cross-sectionally correlated with $\tau=-0.5$. We can observe that, in general, the coverages are only slightly larger than those observed in Figure \ref{fig:Coverages_s} when the idiosyncratic components are cross-sectionally uncorrelated. We can observe nominal coverages around 70\% when $N=25$. Even for large sample sizes, $N=200$ and $T=200$, the coverage is around 90\%. Once more, the undercoverage is associated to the uncertainty in the estimation of the factors, which is larger when there is idiosyncratic cross-sectional correlation; see Fresoli, Poncela and Ruiz (2023, 2025).

\subsection{GARCH idiosyncratic components}

\begin{figure}[h!]
\includegraphics[trim=0 0 0 62, clip, width=1.0\textwidth]{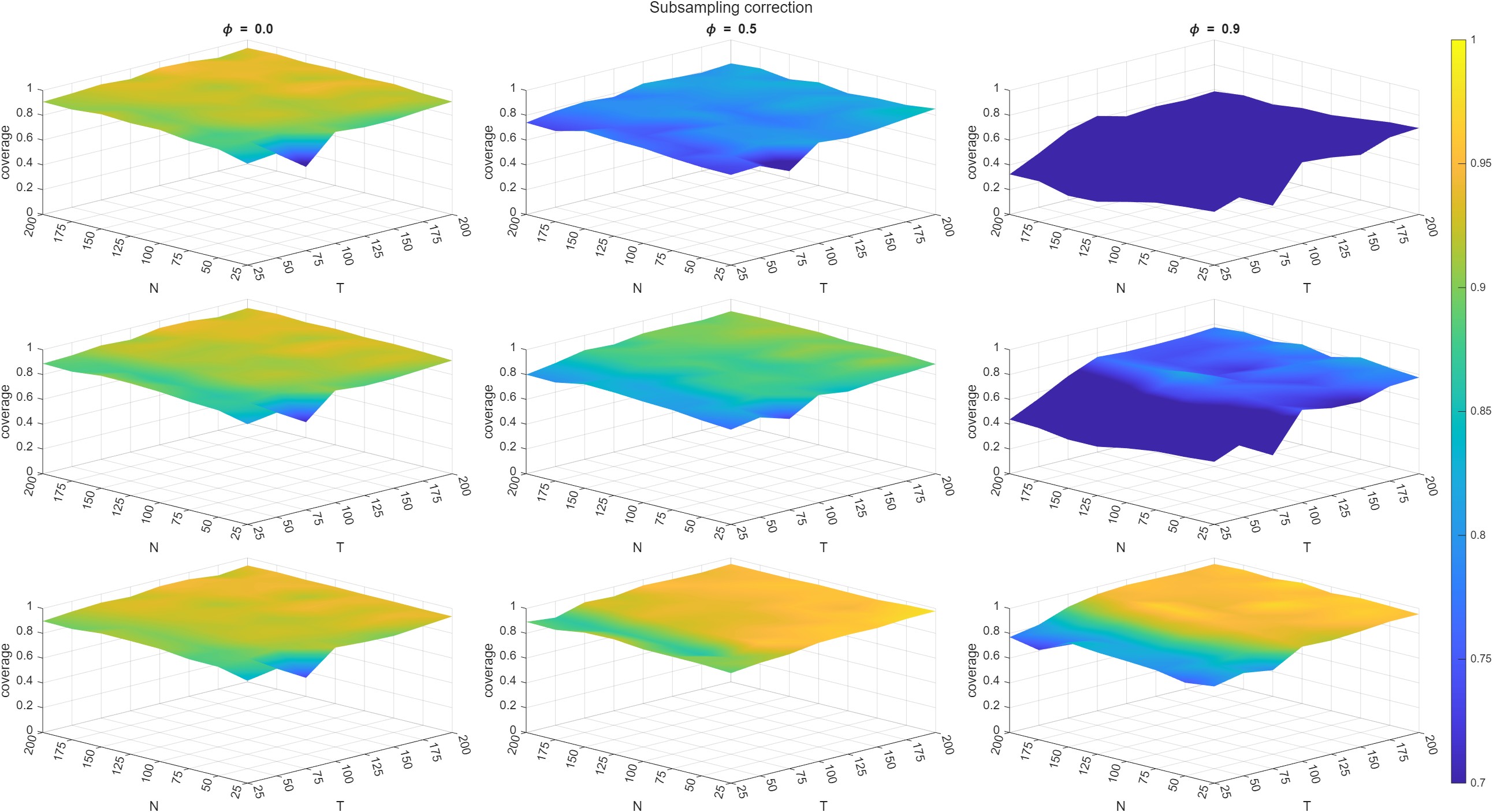}
\vspace{-2mm}
\caption{Empirical coverages of point-wise confidence intervals for $\lambda_1$ for the DFM with $r=1$, $\phi=0.7$ and $c=1$, with idiosyncratic components being cross-sectionally uncorrelated and GARCH(1,1): serially uncorrelated (first column); serially dependent with $\delta=0.5$ (second column); and serially dependent with $\delta=0.9$ (third column). The asymptotic covariance matrix is computed with subsampling correction, without (first row) and with the HAC correction with inference based on normality (second row) and on HAR (third row).}
\label{fig:Coverage_GARCH}
\end{figure}

Figure \ref{fig:Coverage_GARCH} plots the empirical coverages when the DFM has $r=1$ factors generated by an AR(1) model with parameter $\phi=0.7$, the idiosyncratic components are generated by the GARCH(1,1) process defined in equations (\ref{eq:AR}) and (\ref{eq:GARCH}), and the sigmal-to-noise ratio is controlled by $c=1$. When comparing the coverages plotted in Figure \ref{fig:Coverage_GARCH} with those in Figure \ref{fig:Coverages_s}, we can observe that they are very similar, implying that the presence of conditional heteroscedasticity does not affect the results.  

\subsection{Student-t idiosyncratic components}
\begin{figure}[h!]
\includegraphics[trim=0 0 0 62, clip, width=1.0\textwidth]{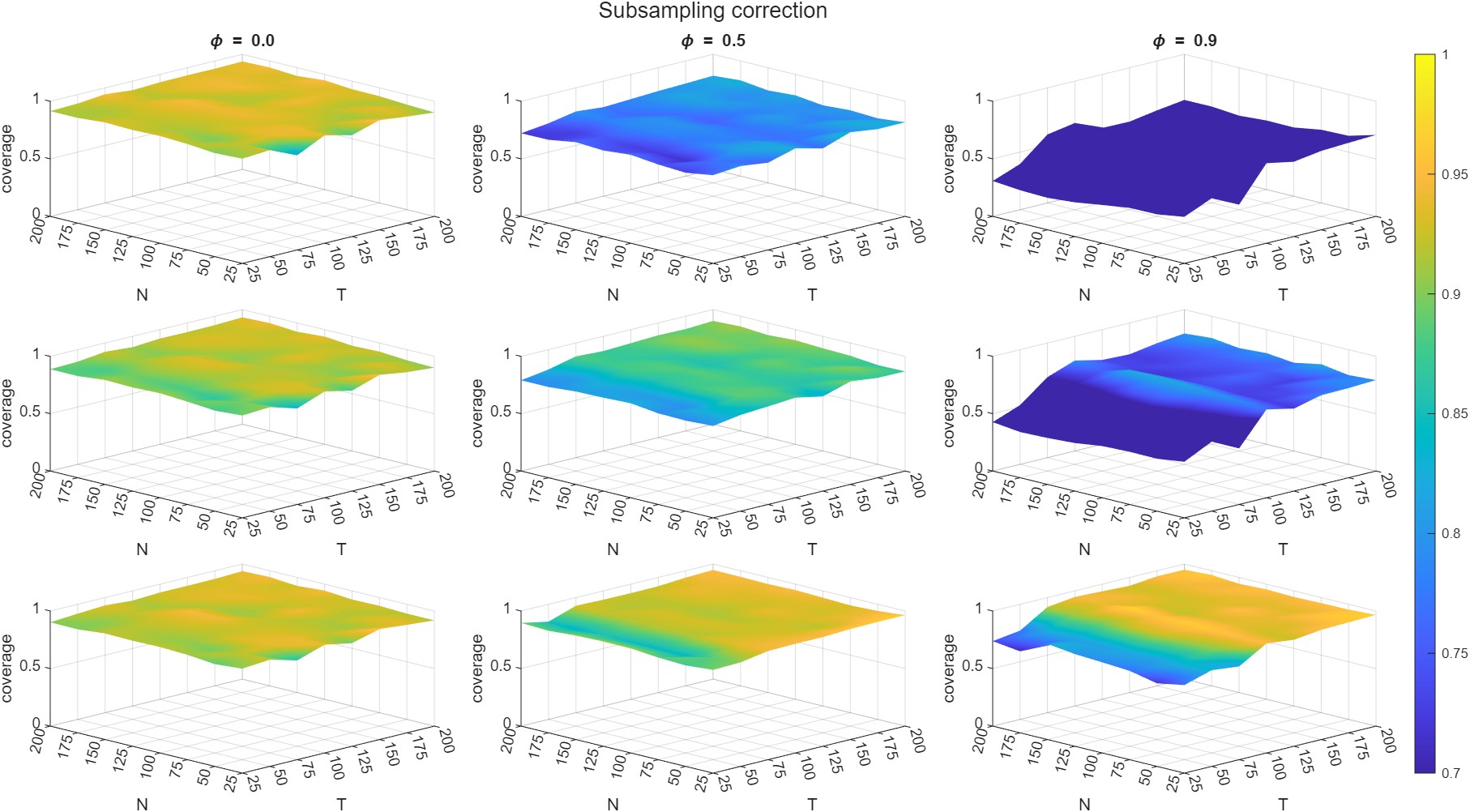}
\vspace{-2mm}
\caption{Empirical coverages of point-wise confidence intervals for $\lambda_1$ for the DFM with $r=1$, $\phi=0.7$ and $c=1$, with idiosyncratic components being cross-sectionally uncorrelated and Student-7: serially uncorrelated (first column); serially dependent with $\delta=0.5$ (second column); and serially dependent with $\delta=0.9$ (third column). The asymptotic covariance matrix is computed with subsampling correction, without (first row) and with the HAC correction with inference based on normality (second row) and on HAR (third row).}
\label{fig:Coverage_Student}
\end{figure}

Figure \ref{fig:Coverage_Student} plots the empirical coverages when the idiosyncratic components are generated by a Student-7 distribution. Once more, we can observe that the coverages are very similar to those in Figure \ref{fig:Coverages_s} when the idiosyncratic components are normally distributed.

\newpage

\section*{Appendix B: Further empirical results}

\setcounter{figure}{0}
\renewcommand{\thefigure}{B.\arabic{figure}}

\setcounter{secnumdepth}{0}

In this Appendix, we report further empirical results on the fit of the DFM to ratios of \textit{per capita} GDP in the $N=51$ states in the US.

\begin{figure}[h!]
\includegraphics[width=0.8\textwidth]{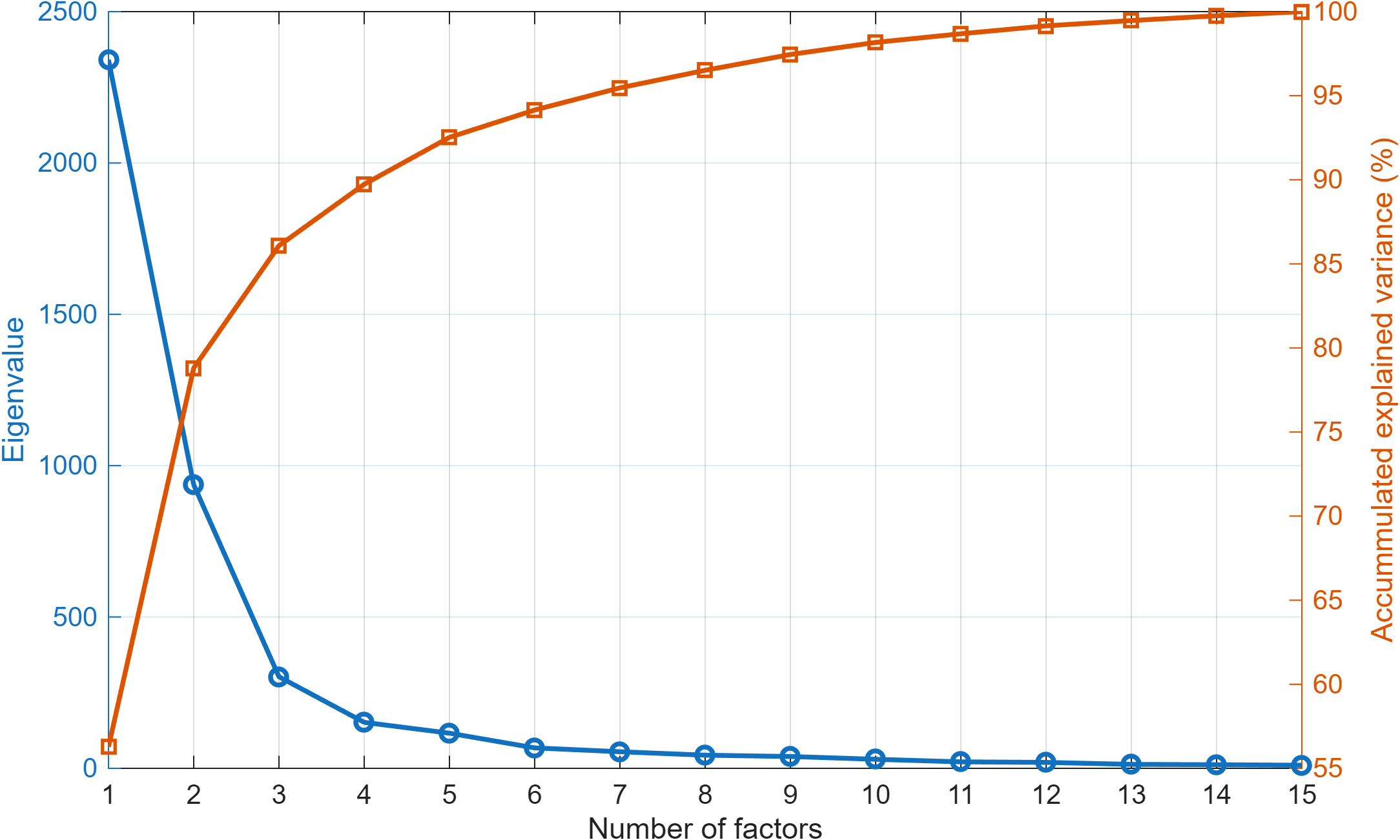}
\caption{Scree plot.}
\label{fig:Scree}
\end{figure}

Figure \ref{fig:Scree} plots the scree-plot corresponding to the series $y_{it}$, $i=1,...,51$, and $t=1,...,84$, together with the cumulative sum of explained variability associated with different number of factors. According to the information in Figure \ref{fig:Scree}, the elbow in the scree-plot is observed for $r=4$ factors, which explain 90\% of the overall variance. 

\begin{figure}[h!]
\includegraphics[width=1.0\textwidth]{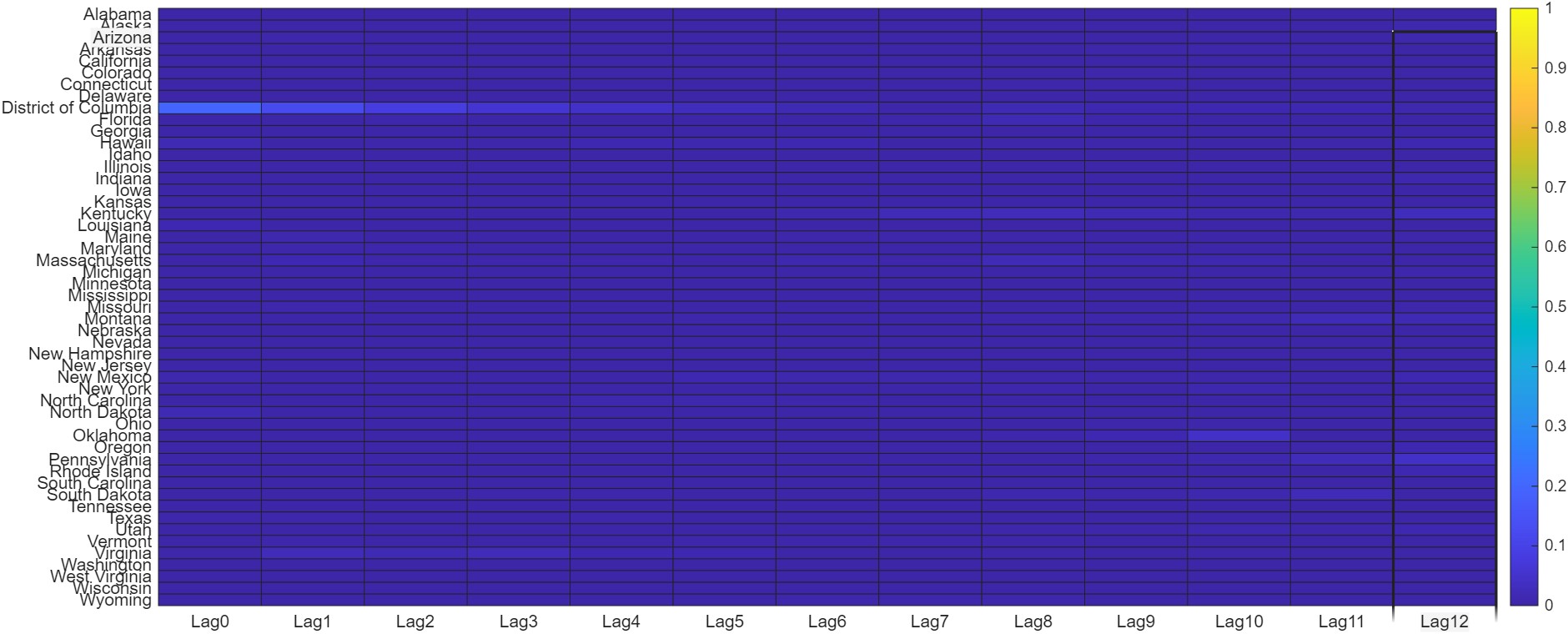}
\caption{$p$-values of ADF test for idiosyncratic residuals in each state for different lags.}
\label{fig:ADF}
\end{figure}

The $p$-values of the ADF test for the idiosyncratic residuals are plotted in Figure \ref{fig:ADF}. This figure shows that the null hypothesis of a unit root is clearly rejected in almost all states.  The only exception is the District of Columbia. 

\begin{figure}[h!]
\includegraphics[width=1.0\textwidth]{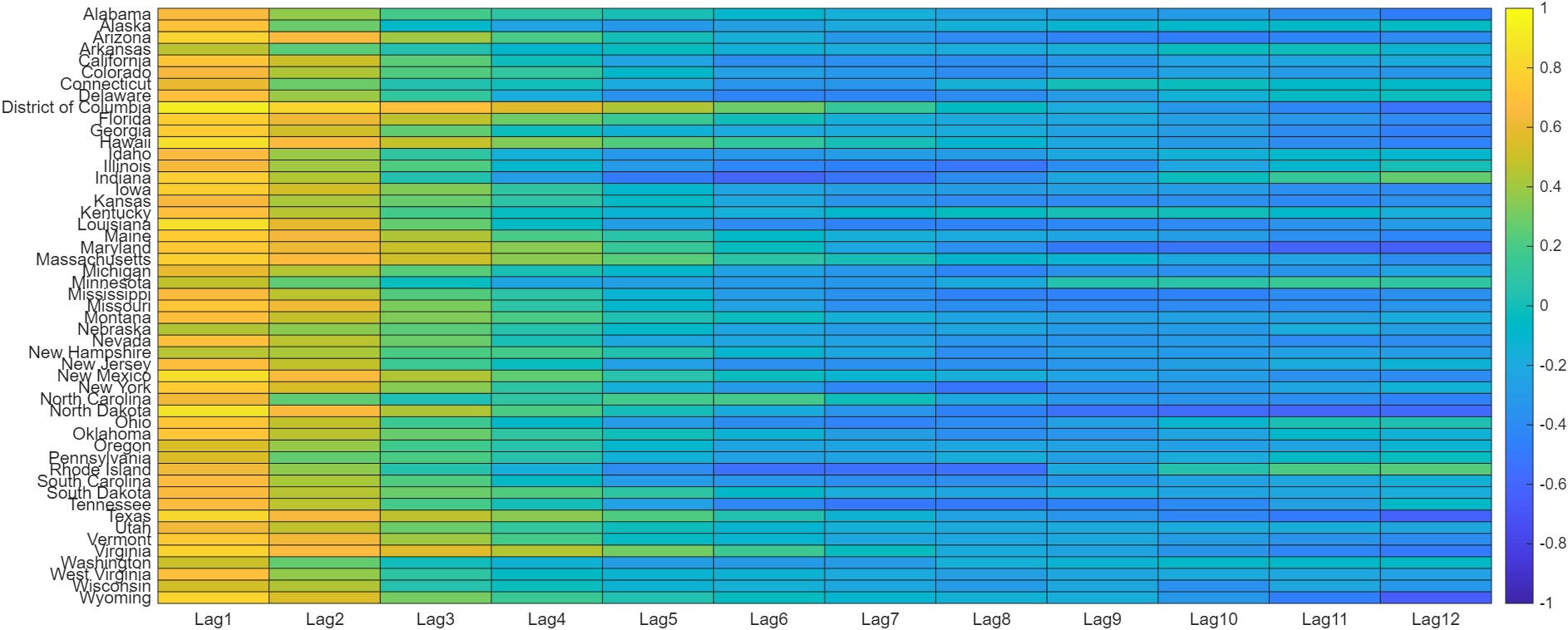}
\caption{Sample autocorrelations of idiosyncratic residuals in each state for different lags.}
\label{fig:correidiosyn}
\end{figure}

Figure \ref{fig:correidiosyn} plots the colormap of the sample autocorrelations of the idiosyncratic residuals. We can observe that, in most states, the temporal dependence of the idiosyncratic residuals can be well represented by AR(1) models with parameters smaller than 0.7.

\begin{figure}[h!]
\includegraphics[trim=0 25 0 25, clip, width=1.0\textwidth]{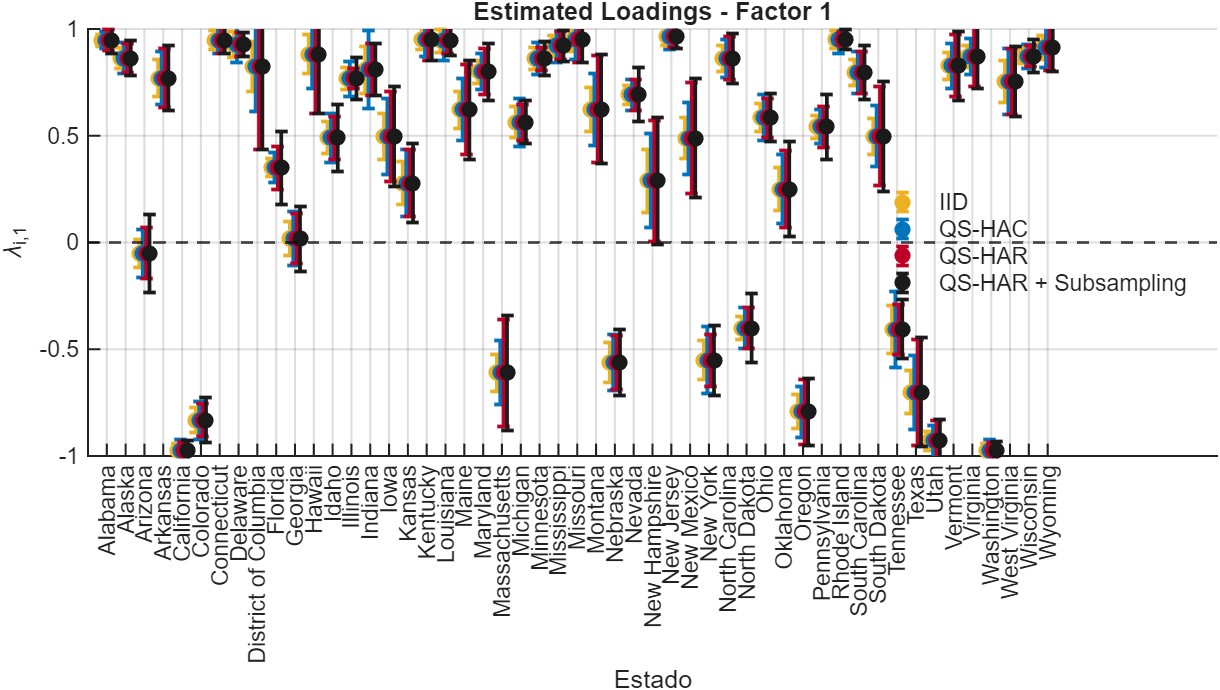}
\caption{Estimated loadings of first factor (black bullet) together with 95\% confidence bounds obtained with normal critical values and the covariance $\Omega$ matrix estimated using: i) $\hat{\Omega}^{(0)}$ (yellow); ii) $\hat{\Omega}^{(QS)}$ (blue), and with HAR critical values and the covariance matrix estimated with $\hat{\Omega}^{(QS)}$ without (red) and with (black) subsampling correction.}
\label{fig:Loadings1_empiric}
\end{figure}

\begin{figure}[h!]
\includegraphics[trim=0 25 0 25, clip, width=1.0\textwidth]{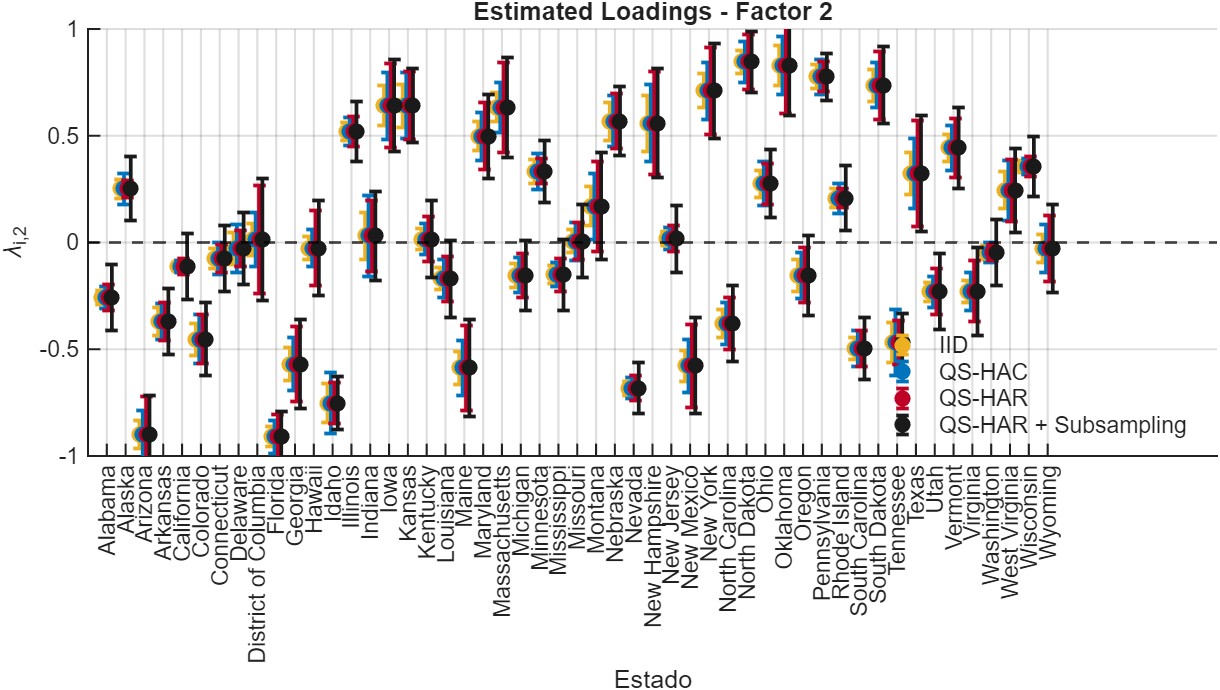}
\caption{Estimated loadings of second factor (black bullet) together with 95\% confidence bounds obtained with normal critical values and the covariance matrix $\Omega$ estimated using: i) $\hat{\Omega}^{(0)}$ (yellow); ii) $\hat{\Omega}^{(QS)}$ (blue), and with HAR critical values and the covariance matrix estimated with $\hat{\Omega}^{(QS)}$ without (red) and with (black) subsampling correction.}
\label{fig:Loadings2_empiric}
\end{figure}

\begin{figure}[h!]
\includegraphics[trim=0 25 0 25, clip, width=1.0\textwidth]{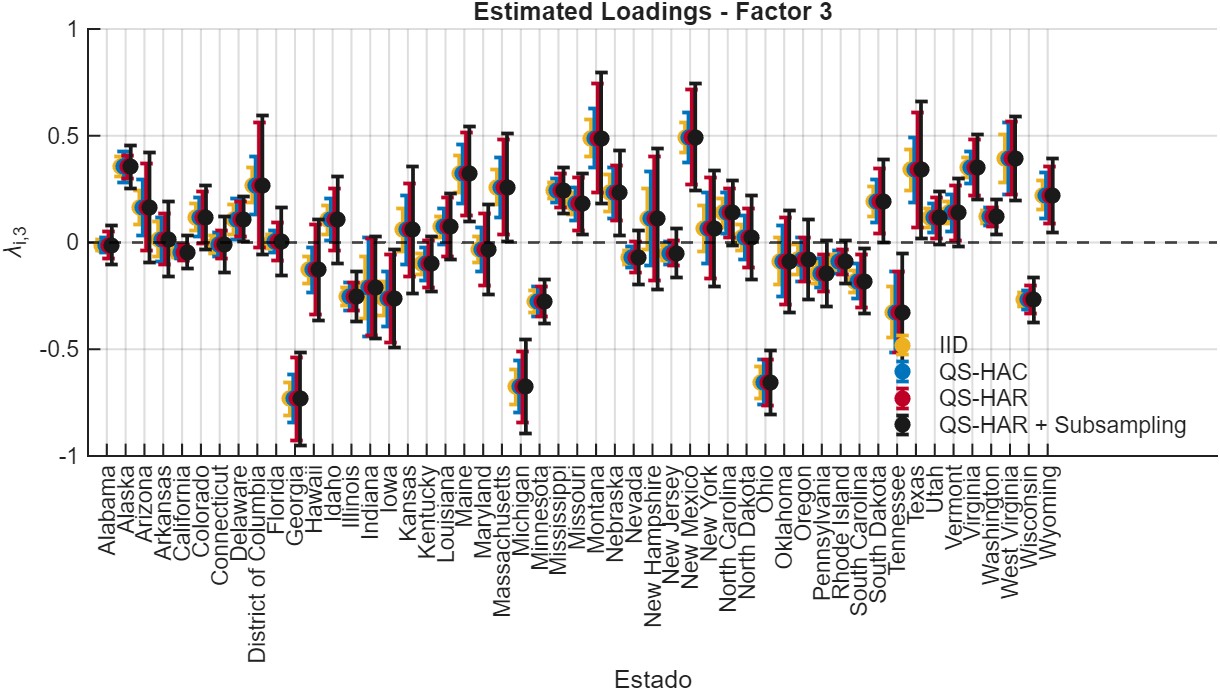}
\caption{Estimated loadings of third factor (black bullet) together with 95\% confidence bounds obtained with normal critical values and the covariance matrix $\Omega$ estimated using: i) $\hat{\Omega}^{(0)}$ (yellow); ii) $\hat{\Omega}^{(QS)}$ (blue), and with HAR critical values and the covariance matrix estimated with $\hat{\Omega}^{(QS)}$ without (red) and with (black) subsampling correction.}
\label{fig:Loadings3_empiric}
\end{figure}

\begin{figure}[h!]
\includegraphics[trim=0 25 0 25, clip, width=1.0\textwidth]{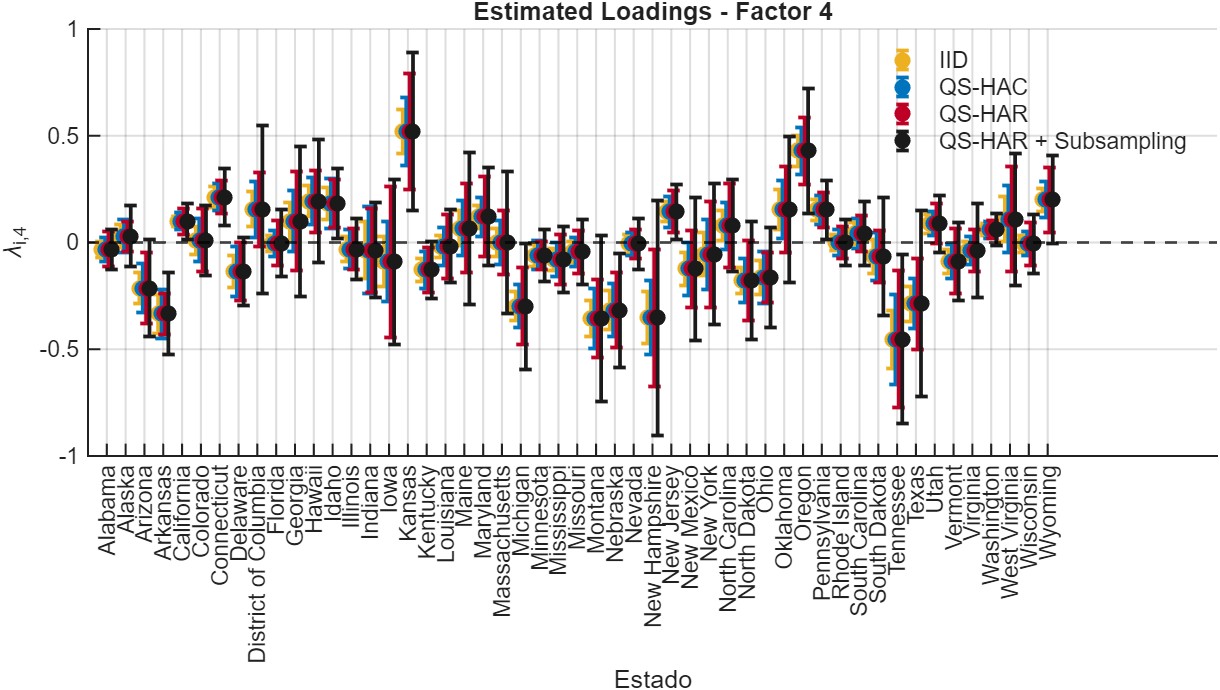}
\caption{Estimated loadings of fourth factor (black bullet) together with 95\% confidence bounds obtained with normal critical values and the covariance matrix $\Omega$ estimated using: i) $\hat{\Omega}^{(0)}$ (yellow); ii) $\hat{\Omega}^{(QS)}$ (blue), and with HAR critical values and the covariance matrix estimated with $\hat{\Omega}^{(QS)}$ without (red) and with (black) subsampling correction.}
\label{fig:Loadings4_empiric}
\end{figure}

Finally, Figures \ref{fig:Loadings1_empiric} to \ref{fig:Loadings4_empiric} plot the estimated PC loadings together with 95\% confidence intervals obtained with the covariance matrix of the loadings estimated by the different procedures considered in this paper. 

\end{document}